\documentclass[pre,aps,twocolumn,preprintnumbers,showpacs,superscriptaddress]{revtex4}
\usepackage{latexsym,epsfig,amssymb,amsfonts,amsmath,graphicx,bbm,mathptm}

\newcommand{\ket}[1]{ | \, #1 \rangle}
\newcommand{\kets}[1]{ | \, #1 \rangle}
\newcommand{\bra}[1]{ \langle #1 \,  |}

\newcommand{\brakets}[2]{\langle\, #1\,|\,#2\,\rangle}

\newcommand{\eqr}[1]{Eq.~(\ref{#1})}
\newcommand{\fir}[1]{Fig.~\ref{#1}}
\newcommand{\secr}[1]{Sec.~(\ref{#1})}
\newcommand{\apr}[1]{Appendix~\ref{#1}}

\begin{document}

\title{Dynamical simulations of classical stochastic systems using matrix product states}

\author{T. H. Johnson}
\email{t.johnson1@physics.ox.ac.uk} 
\affiliation{Clarendon Laboratory, University of Oxford, Parks
Road, Oxford OX1 3PU, United Kingdom}
\author{S. R. Clark} 
\affiliation{Centre for Quantum Technologies, National University of
Singapore, 3 Science Drive 2, Singapore 117543}
\affiliation{Clarendon Laboratory, University of Oxford, Parks
Road, Oxford OX1 3PU, United Kingdom}
\affiliation{Keble College, University of Oxford, Parks
Road, Oxford OX1 3PG, United Kingdom}
\author{D. Jaksch}
\affiliation{Clarendon Laboratory, University of Oxford, Parks
Road, Oxford OX1 3PU, United Kingdom}
\affiliation{Centre for Quantum Technologies, National University of
Singapore, 3 Science Drive 2, Singapore 117543}
\affiliation{Keble College, University of Oxford, Parks
Road, Oxford OX1 3PG, United Kingdom}

\date{\today}

\begin{abstract}
We adapt the time-evolving block decimation (TEBD) algorithm, originally devised to simulate the dynamics of 1D quantum systems, to simulate the time-evolution of non-equilibrium stochastic systems. We describe this method in detail; a system's probability distribution is represented by a matrix product state (MPS) of finite dimension and then its time-evolution is efficiently simulated by repeatedly updating and approximately re-factorizing this representation. We examine the use of MPS as an approximation method, looking at parallels between the interpretations of applying it to quantum state vectors and probability distributions. In the context of stochastic systems we consider two types of factorization for use in the TEBD algorithm: non-negative matrix factorization (NMF), which ensures that the approximate probability distribution is manifestly non-negative, and the singular value decomposition (SVD). Comparing these factorizations we find the accuracy of the SVD to be substantially greater than current NMF algorithms. We then apply TEBD to simulate the totally asymmetric simple exclusion process (TASEP) for systems of up to hundreds of lattice sites in size. Using exact analytic results for the TASEP steady state, we find that TEBD reproduces this state such that the error in calculating expectation values can be made negligible, even when severely compressing the description of the system by restricting the dimension of the MPS to be very small. Out of the steady state we show for specific observables that expectation values converge as the dimension of the MPS is increased to a moderate size.
\end{abstract}


\maketitle

\section{Introduction}
Non-equilibrium stochastic systems have attracted interest for two key reasons. Firstly, they exhibit a wealth of non-trivial phenomena, not observed in equilibrium systems, such as boundary-induced phase transitions~\cite{Krug1991} and infinite range correlations~\cite{Grinstein1990}. Secondly, systems far away from equilibrium have applications in describing a variety of driven diffusive processes, including vehicular traffic flow~\cite{Schadschneider2002,Helbing2001,Chowdhury2000} and biological transport~\cite{Chowdhury2005,Aghababaie1999,Chou1999,Parmeggiani2003}. There exists a range of analytical approaches with which to address such systems, including matrix product methods~\cite{Derrida1993,Blythe2007} and the Bethe ansatz~\cite{Schutz2007}. However, there is no complete analytical framework for non-equilibrium systems, especially away from the steady state, which increases the importance of numerical techniques. 

To simulate these systems, we adapt the time-evolving block decimation (TEBD) algorithm, a well established numerical technique for studying the time-evolution of quantum systems~\cite{Vidal2003,Vidal2004}. TEBD has been used to compute the dynamics of spin chains~\cite{Vidal2004,Gobert2005,White2004} and the Bose Hubbard model~\cite{Clark2004,Daley2004}, which describes an array of 1D quantum systems including cold atoms in optical lattices~\cite{Jaksch2004}. The algorithm provides access to various dynamical properties like atomic many-body currents~\cite{Daley2005}, non-linear responses to driving fields~\cite{Clark2006} and the formation and transport of quasi-particles~\cite{Bruderer2007,Bruderer2008}. The adaption we present here simulates the time-evolution of non-equilibrium stochastic systems, and we refer to it as classical TEBD (cTEBD). 

The principle method used to simulate non-equilibrium systems is dynamic Monte Carlo (DMC), first introduced by~\cite{Gillespie1976}, and two recent examples can be found in~\cite{Popkov2008,Lipowski2009}. cTEBD provides a means of calculating expectation values in a fundamentally different way to DMC simulations. The latter averages over observable values calculated from many trajectories, each a possible evolution of the system, while here we evolve an approximation to the probability distribution of the system from which observable values are calculated. The motivation is to provide a complimentary method to DMC with a contrasting source of error; rather than insufficient sampling we deal with errors arising from approximating a probability distribution by a matrix product state (MPS)~\cite{Fannes1992,Ostlund1995,Rommer1996,PerezGarcia2007}.

Related to TEBD and MPS, a family of numerical techniques originally devised to calculate ground state properties of quantum systems, and later applied to classical non-equilibrium systems, is the density matrix renormalisation group (DMRG) methods~\cite{White1992,White1993,Schollwock2005}. The application of DMRG methods to stochastic non-equilibrium systems in~\cite{Hieida1998,Carlon1999,Carlon2001} suggest that probability distributions of many systems may be approximated well by MPS of small dimension. However, DMRG methods may only be applied to steady states of stochastic systems, not general dynamics which we address in this paper. Also, as shown in~\cite{Carlon1999} for the reaction-diffusion model, they face numerical instabilities for large systems, arising from non-Hermitian diagonalization. 
\begin{figure}[tb]
\includegraphics[width=8cm]{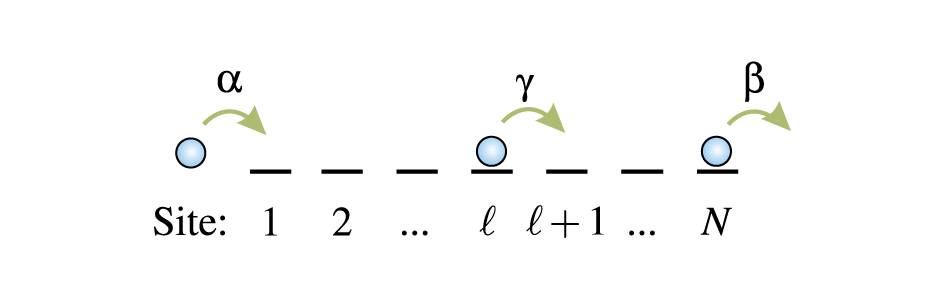}
\caption{(Color online) The TASEP is a paradigm of stochastic non-equilibrium systems. Three Poisson processes contribute to the dynamics of the system. Particles are injected into the first site with rate $\alpha$ if it is empty and ejected from the $N$-th site with rate $\beta$ if occupied. Finally, particles in the $\ell$-th site hop to the $(\ell+1)$-th site with a rate $\gamma$ if that site is empty, resulting in flow from left to right.}\label{fig:ASEP_model}
\end{figure}

A key premise that motivates us to adapt the quantum TEBD algorithm is that a large class of Markovian stochastic systems evolve according to a Schr\"{o}dinger-like equation~\cite{Glauber1963,Alcaraz1993}. In this case the evolution is in imaginary time and governed by a non-Hermitian stochastic Hamiltonian. We demonstrate how this similarity between quantum and stochastic evolution can be exploited by applying cTEBD to the totally asymmetric simple exclusion process (TASEP)~\cite{Golinelli2006,Derrida1998}. The TASEP, shown schematically in \fir{fig:ASEP_model}, is the quintessential driven particle hopping system. Such systems share the following features: particles occupy a 1D lattice; the boundary sites are connected to particle reservoirs, so particles are injected and ejected there; and a hopping process causes particles to move through the lattice (see~\cite{Schmittmann2007,Stinchcombe2001} for reviews). Simple particle hopping systems have drawn much attention, like the Ising and Heisenberg models of equilibrium statistical physics~\cite{Sachdev2001}, because they are archetypal of non-equilibrium physics. Importantly, the TASEP has a Hamiltonian that consists of single-site and two-site nearest-neighbor terms only~\cite{Derrida1993,Rajewsky2004}. Using this, we will outline how the cTEBD algorithm approximately time-evolves a probability distribution by the repeated application of two-site nearest-neighbor operators, each of which affects only the matrices of the MPS associated with those two sites. After each two-site operation the algorithm ensures that the description of the system remains compressed by using a matrix factorization, returning it to an MPS. We have adapted the TEBD to allow for a range of factorizations to be used as part of the algorithm, and choosing which factorization is the key element in determining the computational cost and accuracy of cTEBD. 

The two candidates we focus on are the singular value decomposition (SVD)~\cite{Golub1996} and the non-negative matrix factorization (NMF)~\cite{Paatero1994,Lee1999}, which by restricting the matrices of the MPS to be non-negative (called an sMPS) allows for an information theoretic interpretation, as recently investigated in~\cite{Temme2010}. We compare the performance of these two factorizations and find that while SVD algorithms are well developed, stable and efficient, current NMF algorithms are unable to identify accurate factorizations, due to the non-convexity of the task.

As a result we focus on the SVD-based cTEBD algorithm and investigate its performance in simulating the TASEP. We look in detail at the two main sources of error: the Trotter error, which is due to approximating the evolution over each time-step by a sequence of two-site operations, and the truncation error, due to compressing the state to an MPS of small dimension. For small systems we show how these errors behave, as well as some bounds on them, and then analyze how these results scale with the system size up to hundreds of lattice sites. 
The results obtained promise that cTEBD could be applicable to and accurate in simulating a wide range of non-equilibrium systems.

The structure of the paper is as follows: We review MPS, their construction and physical interpretation in \secr{tebdsec}. Next, in \secr{sec:TEBD} we show how MPS can be evolved efficiently and discuss the properties of two factorization methods for use in cTEBD, namely NMF and SVD. The TASEP and its mathematical representation are introduced in \secr{sec:ASEP}. In \secr{sec:small_sys} we compare the performance of cTEBD with NMF and SVD, before focusing on the errors of the SVD-based algorithm for small systems. We then go on to apply cTEBD to the study of larger systems in \secr{sec:scalability} before presenting our conclusions and outlook for future work in \secr{sec:conclusions}.

\section{Matrix product states}
\label{tebdsec}
Matrix product states are essentially a special structure for approximating vectors in extremely high dimension vector spaces, which typically arise from the tensor product of smaller vector spaces. Since our use of MPS for stochastic systems is motivated by and in many ways analogous to their use in 1D quantum systems, we begin by introducing MPS first for quantum state vectors~\cite{Vidal2003,Verstraete2009,Cirac2009} and then in the context of probability distributions. 

\subsection{Quantum systems}
\label{sec:quantum_mps}
To illustrate MPS in their most general form we consider a quantum system composed of $N$ sites each with a local $d$-dimensional Hilbert space. An arbitrary quantum state of this system $\ket{\psi}$, normalized according to the $L_2$-norm as $\|\ket{\psi}\|_2~=~1$, can be expressed as
\begin{equation}
\ket{\psi} =  \sum_{\bf i} \psi_{\bf i}\ket{\bf i}, \nonumber
\end{equation}
where ${\bf i}=(i_1, i_2,\cdots, i_N)$ is an N-tuple of local indices $i_\ell \nolinebreak =  \nolinebreak 0,\cdots,d-1$ specifying a complete configuration and $\{ \ket{\mathbf{i}} \nolinebreak= \nolinebreak \ket{i_1}\ket{i_2}\cdots\ket{i_N} \}$ are the set of orthonormal configuration states. Since $\ket{\psi}$ inhabits the tensor product vector space $(\mathbbm{C}^d)^{\otimes N}$, whose dimension grows exponentially with $N$ as $d^N$, an exact description of its amplitudes $\psi_{\bf i}$ rapidly becomes intractable with increasing $N$. To attempt to overcome this scaling, we expand $\psi_{\bf i}$ as
\begin{equation}
\psi_{\bf i} = A^{[1] i_1} A^{[2] i_2}\cdots A^{[N] i_N}, \label{eq:mps}
\end{equation}
where for each $i_\ell$ the factor $A^{[\ell] i_\ell}$ is a complex matrix $\mathbbm{C}^{\chi_{\ell-1} \times \chi_{\ell}}$, aside from the boundary terms $A^{[1] i_1}$ and $A^{[N] i_N}$ which are instead complex vectors $\mathbbm{C}^{1 \times \chi_1}$ and $\mathbbm{C}^{\chi_{N-1} \times 1}$, respectively \footnote{It follows from \eqr{eq:mps} that an MPS is invariant under the transformation $A^{[\ell] i_\ell} \mapsto A^{[\ell] i_\ell} X$, $A^{[\ell+1] i_{\ell + 1}} \mapsto X^{-1} A^{[\ell+1] i_{\ell + 1}}$, where $X$ is a non-singular matrix. The formalism of \secr{sec:quantum_mps} and \apr{sec:mps_construct} automatically exploits this freedom so that the tensors $\{ A^{[\ell]} \}$ satisfy orthonormality conditions.}. The expansion in \eqr{eq:mps} is simple to visualize; each site $\ell$, depending on its local configuration $i_\ell$, contributes a matrix $A^{[\ell] i_\ell}$ to a site-ordered multiplication of matrices which results in the amplitude $\psi_{\bf i}$. The boundary vectors then ensure that a scalar is recovered. Such an MPS can be thought of as a tensor network, as depicted in \fir{fig:mps}. A product state, whose MPS requires dimensions $\chi_\ell = 1$ for all $\ell$ has no correlations between different sites, while the MPS describing a vector with arbitrarily strong correlations may need matrix dimensions $\chi_{\ell}$ up to $X_{\ell} \equiv \min \left(d^\ell, d^{N-\ell} \right)$, and so scale exponentially with the system size $N$. Thus in describing arbitrary vectors an MPS description offers no advantage.

The utility of an MPS formulation, however, is based precisely on the fact that physical states have a structure which allows for a near exact MPS description while restricting $\chi_\ell \le \chi$ with $\chi$ much smaller than the largest $X_\ell$. For quantum systems, correlations in a many-body state $\ket{\psi}$ are quantified by the entanglement entropy and this in turn has a rigorous connection to the required MPS dimension $\chi$~\cite{Vidal2003,Verstraete2006}. This is shown by first splitting the system into two contiguous blocks ${\tt L}^{[\ell]} = \{1,\cdots,\ell\}$ and ${\tt R}^{[\ell]}=\{\ell+1,\cdots,N\}$. The state is then expanded as $\ket{\psi} = \sum_{\bf ij} \psi_{\bf ij}\ket{\bf i}_{\tt L}\ket{\bf j}_{\tt R}$, where $\bf i$ and $\bf j$ label configurations of the two subsystems and $\psi$ is now the reshaped $d^\ell \times d^{N-\ell}$ matrix of amplitudes. To expose the correlations of this state an SVD~\cite{Golub1996} is performed which factorizes $\psi = UDV^\dagger$, where $U$ and $V$ are unitary matrices and $D$ is a diagonal matrix with real non-negative elements. This establishes a so-called Schmidt decomposition~\cite{Nielson2000} of $\ket{\psi}$ as
\begin{equation}
\ket{\psi} = \sum_{{\mu} = 1}^{X_\ell} \lambda^{[\ell]}_{{\mu}}\kets{{\tt L}^{[\ell]}_{\mu}}\kets{{\tt R}^{[\ell]}_{\mu}}. \label{eq:schmidt}
\end{equation}
Here $\{ \kets{{\tt L}^{[\ell]}_{\mu}} \}$ and $\{ \kets{{\tt R}^{[\ell]}_{\mu}} \}$ are Schmidt states derived from the columns of $U$ and $V$, while $\{ \lambda^{[\ell]}_{\mu} \}$ are the singular values (Schmidt coefficients in this context) specified by the diagonal elements of $D$, arranged in non-increasing order with $\mu$. The normalization of $\ket{\psi}$ and unitarity of $U$ and $V$ result in
\begin{equation}
\sum_{\mu} (\lambda^{[\ell]}_{\mu})^2 = 1 \quad \textrm{and} \quad \brakets{{\tt L}^{[\ell]}_{\nu}}{{\tt L}^{[\ell]}_{\mu}} = \brakets{{\tt R}^{[\ell]}_{\nu}}{{\tt R}^{[\ell]}_{\mu}} = \delta_{{\nu}{\mu}}. \nonumber
\end{equation}
The mutual information between the subsystems ${\tt L}^{[\ell]}$ and ${\tt R}^{[\ell]}$ is equal to twice the entanglement entropy~\cite{Amico2008}, given by
\begin{equation}
 S_{\tt LR} = -\sum_{{\mu} = 1}^{X_\ell} (\lambda^{[\ell]}_{\mu})^2 \log\left[(\lambda^{[\ell]}_{\mu})^2\right]. \nonumber
\end{equation}
\begin{figure}[tb]
\includegraphics[width=7cm]{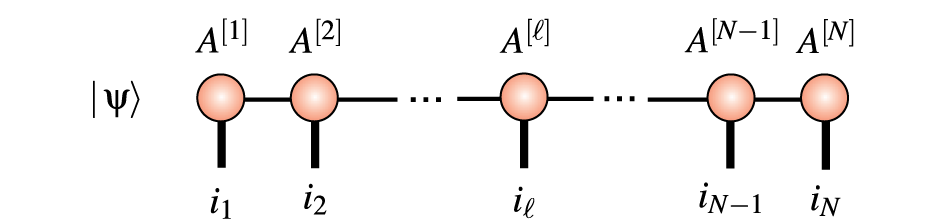}
\caption{(Color online) An MPS for a vector $\ket{\psi}$ as a comb-like tensor network. Each site $\ell$ of the system has a tensor $A^{[\ell]}$ associated with it, illustrated by the shaded circles. The thick vertical leg of a tensor $A^{[\ell]}$ represents the physical index $i_\ell$, while the thinner leg(s) portray the internal index (indices). The boundary tensors have a single internal leg and the others have two. The joining of tensor legs represents the contraction (or matrix multiplication) over the internal indices.}\label{fig:mps}
\end{figure}
The size of the entanglement entropy between a subsystem and the rest of the system for ground states has been extensively studied in the context of area laws (see~\cite{Eisert2010} for a review). These have shown that $S_{\tt LR}$ should scale with the boundary connecting the two parts up to logarithmic corrections at criticality, and area laws therefore severely limit the entanglement entropy physical systems can possess. For quantum spin chains with short-range interactions the area law is manifested in an extremely rapid decay of the Schmidt coefficients $\{ \lambda^{[\ell]}_{\mu} \}$~\cite{Vidal2004}.  An immediate consequence of this is that an accurate (un-normalized) approximation $\ket{\tilde{\psi}^{[\ell]}}$ of $\ket{\psi}$ can be formed by truncating the sum in \eqr{eq:schmidt} to the first $\chi$ terms~\cite{Vidal2004} as
\begin{equation}
\ket{\tilde{\psi}^{[\ell]}} = \sum_{{\mu} = 1}^{\chi} \lambda^{[\ell]}_{{\mu}}\kets{{\tt L}^{[\ell]}_{\mu}}\kets{{\tt R}^{[\ell]}_{\mu}}. \label{eq:schmidt_trunc} \nonumber
\end{equation}
The error made in computing the expectation value of any quantum mechanical observable with $\ket{\tilde{\psi}^{[\ell]}}$ rather than $\ket{\psi}$ is bounded by the $L_2$-norm of the residual~\cite{Nielson2000}, given by $||\ket{\psi} - \ket{\tilde{\psi}^{[\ell]}}||_2 = [\sum_{{\mu}=\chi+1}^{X_\ell} (\lambda^{[\ell]}_{\mu})^2]^{1/2}$ and is therefore small if the sum of the squares of the truncated Schmidt coefficients is small~\cite{Vidal2004}. These observations can be directly related to an MPS, as described in \apr{sec:mps_construct}, by applying this decomposition and truncation in an iterative sequence to all contiguous bipartitions to form an approximation $\ket{\tilde{\psi}}$~\cite{Vidal2003}. This approximation is an MPS of the form of \eqr{eq:mps} but with a maximum matrix dimension $\chi$. The errors in this case are bounded by the singular values~\cite{Verstraete2006} as
\begin{equation}\label{eq:quanterror}
||\ket{\psi} - \ket{\tilde{\psi}}||_2 \le \left[ 2 \sum_{\ell = 1}^{N-1} \sum_{{\mu}=\chi+1}^{X_\ell} (\lambda^{[\ell]}_{\mu})^2 \right]^{1/2} .
\end{equation}
A rapid decay of $\{ \lambda^{[\ell]}_\mu \}$ with $\mu$ for each bipartition thus enables $\chi$ to be kept small while retaining a high accuracy, and results in an MPS description parameterized by $O(Nd\chi^2)$ complex numbers~\cite{Vidal2004}. This near lossless compression of the information in $\{ \psi_{\bf i} \}$ for physically relevant vectors is responsible for the enormous success of the MPS approach used within DMRG \cite{White1992,White1993} and TEBD methods \cite{Vidal2003,Vidal2004}.
 
\subsection{Classical systems} 
\label{sec:classical_mps}
The general MPS formalism introduced can equally be applied to stochastic classical systems. We now consider a system of $N$ sites each with $d$ local configurations. The state of the system is then described by a probability vector $\ket{P} \nolinebreak = \nolinebreak \sum_{\bf i} P_{\bf i}\ket{\bf i}$ with $d^N$ real non-negative components $\{ P_{\bf i} \}$ corresponding to the probabilities of the system being in each of the configurations $\{ \bf i \}$. Classical states $\ket{P}$ are therefore contained in the positive orthant of the vector space $(\mathbbm{R}^d)^{\otimes N}$ and are normalized according to the $L_1$-norm as $\|\ket{P}\|_1=1$. An immediate restriction to real matrices $\{ A^{[\ell] i_\ell} \}$ within the MPS expansion in \eqr{eq:mps} can be made in this case. A major question which we will address in this work is whether it is necessary, or indeed desirable, to further restrict the matrices to be non-negative. To do so manifestly ensures that the sMPS approximation to $\ket{P}$ never leaves the positive orthant. Early exact analytical calculations based on an MPS formalism were sMPS~\cite{Derrida1993}.

For classical systems a similar link between the dimension $\chi$ required and correlations in $\ket{P}$ can be established by working exclusively with non-negative quantities. As before, we bipartition the system after site $\ell$ into two contiguous blocks ${\tt L}^{[\ell]}$ and ${\tt R}^{[\ell]}$ and then expand $\ket{P} = \sum_{\bf ij} P_{\bf ij}\ket{\bf i}_{\tt L}\ket{\bf j}_{\tt R}$. Following Temme and Verstraete in~\cite{Temme2010}, we may draw a close analogy to the Schmidt form of a quantum state via a decomposition of $P$ as
 \begin{equation}
P_{\bf ij} = \sum_{{\mu}=1}^{X_\ell}p^{[\ell]}_{\mu} P^{[\ell]}_{\tt L}({\bf i}|{\mu}) P^{[\ell]}_{\tt R}({\bf j}|{\mu}), \label{eq:probschmidt}
\end{equation}
which has the elegant interpretation of having $\{ p^{[\ell]}_{\mu} \}$ as probabilities, again arranged in non-increasing order with $\mu$, which mix together the conditional marginal probability distributions $P^{[\ell]}_{\tt L}$ and $P^{[\ell]}_{\tt R}$. This could equally be expressed as the vector decomposition
 \begin{equation}
\ket{P} = \sum_{{\mu} = 1}^{X_\ell} p^{[\ell]}_{\mu}\kets{{\tt L}^{[\ell]}_{\mu}}\kets{{\tt R}^{[\ell]}_{\mu}}, \label{eq:probschmidtvec}
\end{equation}
where $\kets{{\tt L}^{[\ell]}_{\mu}}$ and $\kets{{\tt R}^{[\ell]}_{\mu}}$ have elements $\{ P^{[\ell]}_{\tt L}({\bf i}|{\mu}) \}$ and $\{ P^{[\ell]}_{\tt R}({\bf j}|{\mu}) \}$ respectively, similar in form to the Schmidt decomposition in \eqr{eq:schmidt} except that $\ket{P}$ is expanded in terms of non-negative vectors rather than orthonormal ones. In this case the mutual information is upper-bounded by the entropy of the mixing probability distribution~\cite{Temme2010}, given by
\begin{equation}
 S(\{p^{[\ell]}_{\mu}\}) = -\sum_{\mu = 1}^{X_\ell} p^{[\ell]}_{\mu} \log\left[p^{[\ell]}_{\mu}\right]. \nonumber
\end{equation}
Akin to the truncation of a Schmidt decomposition, an (un-normalized) approximation $\ket{\tilde{P}^{[\ell]}}$ to $\ket{P}$ can be formed by truncating the decomposition in \eqr{eq:probschmidt} to the first $\chi$ terms, while the $L_1$-norm of the residual is $||\ket{P} \nolinebreak - \nolinebreak \ket{\tilde{P}^{[\ell]}}||_1 \nolinebreak = \nolinebreak \sum_{{\mu}=\chi+1}^{X_\ell} p^{[\ell]}_{\mu}$. In \apr{sec:obs_errors} we show that for classical systems the error made in computing any observable is upper-bounded by the $L_1$-norm error of the probability vector, analogous to the role of the $L_2$-distance for quantum systems. The decomposition \eqr{eq:probschmidt} is in general non-unique and the most significant decomposition at this bipartition is the one with the spectrum $\{p^{[\ell]}_{\mu}\}$ that has the smallest entropy $S_C$, called the entropy cost~\cite{Temme2010}. This suggests that the entropy cost $S_C$ should be considered analogous to the entanglement entropy $S_{\tt LR}$ since it quantifies both the quality of approximation and correlations. However, since $S_C$ only upper-bounds the mutual information the connection between correlations and truncation errors is substantially weaker than that seen in the quantum case. Just like for quantum systems, repeated application of the decomposition in \eqr{eq:probschmidtvec} at all bipartitions can be used to build an approximate sMPS $\ket{\tilde{P}}$, and in~\cite{Temme2010} it was shown that the error is bounded by
\begin{equation}
||\ket{P} - \ket{\tilde{P}}||_1 \le \sum_{\ell = 1}^{N-1} \sum_{{\mu}=\chi+1}^{X_\ell} p^{[\ell]}_{\mu} . \nonumber
\end{equation}
Exact sMPS solutions for TASEP stationary states find that the entropy minimizing spectrum decays very quickly~\cite{Temme2010} thus indicating in a similar way to the quantum case that a small $\chi$ can be chosen while maintaining an accurate description.

In this work we note that each decomposition \eqr{eq:probschmidt} is in fact precisely equivalent to an exact NMF of the matrix $P$, analogous to performing an SVD in the quantum case. The exact NMF can be written as $P = UDV^T$ where the columns of $U$ and $V$ are the normalized non-negative probability vectors in the decomposition \eqr{eq:probschmidtvec} and the diagonal elements of $D$ are the normalized probabilities $\{ p^{[\ell]}_{\mu} \}$~\cite{Gaussier2005}. As was done with the SVD, $\ket{\tilde{P}^{[\ell]}}$ is formed by truncating the inner-dimension of the NMF to $\chi$.

While the NMF may be a more natural choice for decomposing probability vectors at a formal level, it is perfectly possible to employ SVDs on $\ket{P}$ in precisely the same way as for a quantum state. The $L_2$-norm error is given by \eqr{eq:quanterror}, and the $L_2$-norm upper-bounds the more relevant $L_1$-norm with a factor of the square root of the dimension of the vector space~\cite{Golub1996}, hence for the approximation $\ket{\tilde{P}}$ built in this way
\begin{equation}\label{eq:Lambda}
||\ket{P} - \ket{\tilde{P}}||_1 \le \Lambda \equiv 2^{N/2} \left[ 2 \sum_{\ell=1}^{N-1}\sum_{{\mu}=\chi+1}^{X_\ell} (\lambda^{[\ell]}_{\mu})^2 \right]^{1/2}.
\end{equation}
The SVD approach has two pleasing features: firstly, it is unique, and secondly, given $\ket{P} = \sum_{\bf ij} P_{\bf ij}\ket{\bf i}_{\tt L}\ket{\bf j}_{\tt R}$ and $\ket{Q} \nolinebreak = \nolinebreak \sum_{\bf ij} Q_{\bf ij}\ket{\bf i}_{\tt L}\ket{\bf j}_{\tt R}$, a well known result first proposed by Eckart and Young~\cite{Eckart1936} is that $\ket{\tilde{P}^{[\ell]}}$ is the solution to the optimization problem~\cite{Golub1996}
\begin{equation}
\min_{\substack{Q \in \mathbbm{R}^{d^\ell \times d^{N - \ell}}, \\ \textrm{rank}(Q) \le \chi} } || \ket{P} - \ket{Q} ||_2 = || \ket{P} - \ket{\tilde{P}^{[\ell]}} ||_2 \label{eq:eckartyoung}
\end{equation}
for any $1 \leq \chi \leq X_\ell$. Thus the $L_2$-norm error of a truncated SVD will always be less than that of a truncated NMF or any other factorization of the same rank. This ensures that if there exists an exact MPS of some dimension $\chi$ then there also exists an exact SVD-MPS of dimension $\chi$ or less. 

A powerful feature of the SVD is the orthogonality constraint on the singular vectors (the columns of $U$ and $V$). However, this typically results in the $U$ and $V$ matrices having elements of mixed signs, even for non-negative $P$, and so does not permit any obvious information theoretic interpretation of the components of the decomposition, as exists for an NMF. This also means that when $\ket{P}$ is truncated to form $\ket{\tilde{P}^{[\ell]}}$, negative probabilities may enter our description. The maximum negativity of any element of $\ket{\tilde{P}^{[\ell]}}$ is
\begin{eqnarray}
\max_{\bf i,j} |-\tilde{P}^{[\ell]}_{\bf ij}| \le \|\ket{P}-\ket{\tilde{P}^{[\ell]}}\|_\infty &\leq&  \sum_{\mu=\chi+1}^{X_\ell} \lambda^{[\ell]}_{\mu}. \nonumber
\end{eqnarray}
Thus while non-negativity is not conserved it is controlled by a similar criterion to the accuracy of the factorization, namely the smallness of the truncated singular values. This bound is loose, for example in the worst case limit of truncation to $\chi=1$ non-negativity is in fact preserved due to the Perron-Frobenius theorem~\cite{Seneta2006} which ensures that both the left and right singular vectors associated to the largest singular value of a non-negative matrix are themselves non-negative.

\section{Time-evolving block decimation}
\label{sec:TEBD}
We have seen that MPS can provide an accurate and efficient description of some probability distributions. Our purpose for this section is to review the TEBD algorithm, originally devised to update a quantum MPS under unitary time-evolution~\cite{Vidal2003,Vidal2004,Daley2004}, but now paying specific attention to issues arising from its use on classical stochastic systems. Analogous to quantum systems, the evolution of a large class of stochastic systems, for a time $t$, is performed by $\ket{P(t)} \nolinebreak = \nolinebreak \exp(H t)\ket{P}$~\cite{Glauber1963,Alcaraz1993}. The evolution in this case is stochastic and $H$ is a stochastic Hamiltonian, the properties of which will be discussed in \secr{sec:ASEP}. 
\subsection{Time-evolving the state}
\label{sec:trotter}
For anything but the smallest systems neither $\ket{P}$ nor the full stochastic evolution operator $\exp(H t)$ can be represented exactly. Having applied an MPS to tackle the description of $\ket{P}$ we now follow the standard approach used for similar quantum evolutions~\cite{Vidal2004}, namely breaking it up into $n$ small time-steps $\delta t = t/n$ as $\exp(H t) \nolinebreak = \nolinebreak \prod_{j=1}^n \exp(H \delta t)$ and then splitting $\exp(H \delta t)$ into a product of two-site operators. To do the latter we first restrict ourselves to considering stochastic Hamiltonians $H$ composed of a sum of nearest-neighbor two-site terms $\{ h_{\ell,\ell+1} \}$ as
\begin{equation}
H = \sum_{\ell=1}^{N-1} h_{\ell,\ell+1}, \nonumber \label{eq:hamiltonian}
\end{equation}
where we have incorporated any single-site terms into $\{ h_{\ell,\ell+1} \}$. The TASEP is in this restricted class~\cite{Derrida1993,Rajewsky2004}. We can proceed to split up $\exp(H \delta t)$ via a second-order Suzuki-Trotter expansion~\cite{Suzuki1990} as
\begin{equation}
\mathrm{e}^{H \delta t} = \left(\prod_{\ell =1}^{N-1} \mathrm{e}^{h_{\ell,\ell+1} \delta t/2}\right)\left(\prod_{\ell =N-1}^{1} \mathrm{e}^{h_{\ell,\ell+1} \delta t/2}\right) + O\left(\delta t^3 \right), \label{eq:Trotter}
\end{equation}
which represents an approximation because overlapping terms, e.g. $h_{\ell,\ell+1}$ and $h_{\ell+1,\ell+2}$, do not generally commute. This expansion therefore applies a sequence of two-site nearest-neighbor stochastic operators $S_\ell =\exp( h_{\ell,\ell+1}\delta t /2)$ sweeping across pairs of sites, as depicted in \fir{fig:trotter_twosite}(a). A consequence of this expansion is that the approximation of $\exp(H t)$ remains stochastic and so preserves the $L_1$-norm and non-negativity of a probability vector $\ket{P}$. In \apr{sec:trotter_trunc_errors} we show that for stochastic systems, the corresponding $L_1$-norm error incurred by using this expansion, which we call the Trotter error, scales at worst as $\mathcal{E}_{\textrm{ST}} \sim t\delta t^2$.
\begin{figure}[tb]
\includegraphics[width=8.6cm]{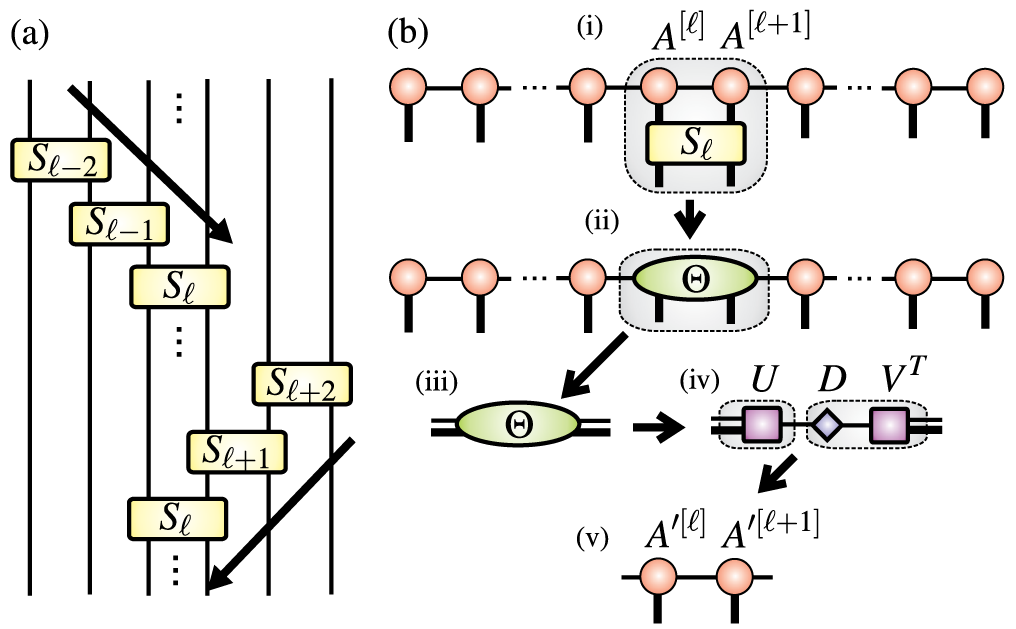}
\caption{(Color online) (a) A circuit diagram showing the second-order Suzuki-Trotter expansion used in this work. One time-step $\exp(H \delta t)$ is approximated by applying two-site stochastic operators $S_\ell$ between each nearest-neighbor pair of sites, sweeping first from left to right and then back again. (b) A schematic tensor diagram of the core procedure in the TEBD algorithm \cite{Vidal2003,Vidal2004} for applying a two-site operator, described in the text. Firstly, in (i) the physical legs of operator $S_\ell$ are contracted with the relevant legs of the tensors $\{ A^{[\ell]}\}$ in the MPS. This multiplies the state by $S_\ell$ and involves only the tensors shaded. The result of this is a two-site order-4 tensor $\Theta$ shown in (ii). Next in (iii) the $\Theta$ tensor is reshaped as a matrix, resulting in two so-called fat legs. The most computationally expensive step is (iv) where an approximate matrix factorization of $\Theta$, of the form $\tilde{\Theta} = UDV^T$ is computed. Finally in (v) the reshaping of $U$ and $DV^T$ is equivalent to unravelling their legs to form the newly updated $A'^{[\ell]}$ and $A'^{[\ell+1]}$ tensors. The diagonal matrix $D$ is absorbed into the $V^T$ during the left to right sweep, as shown by the shading in (iv), and into the $U$ for the opposite direction. The $A'^{[\ell]}$ tensor constructed from an orthogonal matrix only, automatically obeys one of the orthonormality constraints detailed in \apr{sec:mps_construct}. Which way the $D$ is absorbed then follows from requiring that the $\Theta$ tensor for the next two-site operation has indices which correspond to orthonormal bases, as discussed in \secr{sec:optimality}.}\label{fig:trotter_twosite}
\end{figure}  
  
\subsection{Applying a two-site stochastic operator}
\label{sec:two-site}
The core of the TEBD algorithm is a procedure for applying a two-site nearest-neighbor operator to an MPS and performing a systematic truncation of the result back into an MPS of dimension $\chi$~\cite{Vidal2003,Vidal2004}. Using this procedure then allows the sequence of operators $S_\ell$ making up the Trotterized approximation to $\exp(H \delta t)$ to be approximately applied to $\ket{P}$. Given an MPS of form $\ket{P} = \sum_{\bf i} A^{[1] i_1} A^{[2] i_2}\cdots A^{[N] i_N}\ket{\bf i}$ the operator $S_\ell$ can be applied exactly to $\ket{P}$ and only affects the $A^{[\ell]}$ and $A^{[\ell+1]}$ tensors, as shown in step (i) of \fir{fig:trotter_twosite}(b). This forms a new order-4 tensor associated to these sites with the contraction written out explicitly as
\begin{equation}
\Theta^{i_\ell i_{\ell+1}}_{\nu\mu} = \sum_{j_\ell,j_{\ell+1}=0}^{d-1} (S_\ell)^{i_\ell i_{\ell+1}}_{j_\ell j_{\ell+1}} \sum_{\zeta=1}^{\chi_\ell} A^{[\ell] j_\ell}_{\nu\zeta}A^{[\ell+1] j_{\ell+1}}_{\zeta\mu}. \nonumber
\end{equation}
The resulting state $S_\ell\ket{P}$, depicted in step (ii) of \fir{fig:trotter_twosite}(b), no longer has a proper MPS form due to the presence of a structureless two-site $\Theta$ tensor.  To establish an MPS form we first reshape the $\Theta$ tensor into a conventional matrix by combining the indices $(i_\ell,\nu)$ and $(i_{\ell+1},\mu)$ into so-called fat indices which represent the row and columns respectively, as depicted in step (iii) of \fir{fig:trotter_twosite}(b). The resulting $\Theta$ matrix is then subject to a matrix factorization which approximately decomposes it into the form $\tilde{\Theta} = UDV^T$, where $D$ is a $\mathbbm{R}^{\chi \times \chi}$ diagonal matrix, $U \in \mathbbm{R}^{d \chi \times \chi}, V \in \mathbbm{R}^{d \chi \times \chi}$ and $\chi$ is the desired MPS dimension. This is step (iv) in \fir{fig:trotter_twosite}(b) and is the most computationally expensive part of the procedure, requiring a number of computations $O(d^3 \chi^3)$ for the SVD. After absorbing $D$ into either the $U$ or $V^T$ matrix according to the direction of the Trotter sweep, the matrices are reshaped into order-3 tensors $A'^{[\ell]}$ and $A'^{[\ell+1]}$. In this way, as shown in step (v) of \fir{fig:trotter_twosite}(b), the factorization is used to impose internal structure on the original $\Theta$ tensor and by replacing $\Theta$ with the new tensors establishes a proper MPS for $S_\ell\ket{p}$. Crucially, however, the factorization also provides the means of truncating the MPS dimension to some desired maximum $\chi$, controlling the growth of the description at the expense of becoming an approximation. We call the $L_1$-norm error due to these approximate factorizations the truncation error $\mathcal{E}_{\textrm{tr}}$ and we discuss the calculation of quantities which bound this error in \apr{sec:trotter_trunc_errors}. In \secr{tebdsec} we have studied the analytic properties of the SVD and NMF factorizations, while now we focus on the numerical properties of these factorizations and their accuracy when used in the two-site procedure described above.

\subsubsection{Non-negative matrix factorization}
\label{sec:NMFdets}
There is no known algorithm for finding (non-trivial) exact NMF solutions for a given $m \times n$ non-negative matrix $\Theta$, despite their existence~\cite{Vasiloglou2009}, and so an NMF based approximation cannot currently proceed by truncation from an exact factorization. Instead an approximation $\tilde{\Theta} = WH^T$ is formed, where $W$ and $H$ are non-negative matrices, by solving directly the reduced rank minimization problem
\begin{equation}
\min_{\substack{W \in \mathbbm{R}_+^{m \times \chi}, \\ H \in \mathbbm{R}_+^{n \times \chi}}}  \left\{F(\Theta,WH^T)\right\}, \nonumber
\end{equation}
where $F(\Theta,WH^T)$ is a cost function that quantifies the quality of the approximation to $\Theta$. By forming diagonal matrices $D_W$ and $D_H$ containing the column sums of $W$ and $H$ any NMF can be brought into the form $\tilde{\Theta} = UDV^T$ where $D$ is a diagonal matrix $D = D_WD_H$, satisfying $\textrm{tr}(D) = \sum_{\bf{ij}}\Theta_{\bf{ij}}$, and $U$ and $V$ are column stochastic~\cite{Gaussier2005}. Common choices for $F$ are
\begin{equation}
F(\Theta,WH^T) = \left\{ \begin{array}{l@{\,}l} \sqrt{\sum_{\bf{ij}}|\Theta_{\bf{ij}} - (WH^T)_{\bf{ij}}|^2}, \\
\sum_{\bf{ij}}\left\{\Theta_{\bf{ij}}\log\left[\frac{\Theta_{\bf{ij}}}{(WH^T)_{\bf{ij}}}\right] - \Theta_{\bf{ij}} + (WH^T)_{\bf{ij}}\right\}, \\
\sum_{\bf{ij}} |\Theta_{\bf{ij}} - (WH^T)_{\bf{ij}}|,\end{array}\right. \label{eq:costfuncs}
\end{equation}
which are the $L_2$-norm of the residual reshaped as a vector $\|\ket{\Theta}-\ket{\tilde{\Theta}}\|_2$, the generalized Kullback-Leibler (KL) divergence \footnote{Despite not being a distance measure the KL divergence does have an information theoretic interpretation. When $\sum_{ij}(WH^T)_{ij} = 1$, in addition to $\Theta$, the KL divergence reduces to the relative entropy of the two probability distributions. It then quantifies the expected number of extra bits required to code samples from $\Theta$ when using a code based on $WH^T$, rather than using a code based on $\Theta$.}, and the $L_1$-norm $\|\ket{\Theta}-\ket{\tilde{\Theta}}\|_1$, respectively~\cite{Lee2000}. A crucial issue for any minimization algorithm based on these cost functions is that although they are convex for $W$ and $H$ separately they are not convex for $W$ and $H$ simultaneously, meaning the problem can have many local minima. A common strategy for minimization is to alternate the optimization between $W$ and $H$ while the other is fixed, however such algorithms can, at best, only guarantee convergence to a local minimum. Furthermore there is no guarantee that the global minimum, if located, is unique. The non-uniqueness of an NMF is evident given that any $\mathbbm{R}_{+}^{\chi \times \chi}$ non-negative monomial matrix $M$ can also form an alternative NMF as $\tilde{\Theta} = WM M^{-1}H^T$. In \apr{sec:nmf_algorithms} we outline some popular and simple NMF algorithms. From numerical test we have found, as expected, that the $L_1$-norm minimizing NMF algorithms consistently deliver the smallest $L_1$-norm errors and so we concentrate on this cost function.

\subsubsection{Singular value decomposition}  
For any real $m \times n$  matrix $\Theta$ there exists an SVD of the form $\Theta = U D V^T$, where $U$ and $V$ are $\mathbbm{R}^{m \times X}$ and $\mathbbm{R}^{n \times X}$ orthogonal matrices respectively, and $X = \min(m,n)$ is the maximum possible rank of $\Theta$~\cite{Golub1996}. In this context $D$ is a $\mathbbm{R}^{X \times X}$ diagonal matrix of non-negative singular values $\{ \lambda_\mu \}$ arranged in non-increasing order with $\mu$, and the columns of $U$ and $V$ are the singular vectors. In contrast to the NMF, the SVD is a unique decomposition up to the signs of its left and right singular vectors. These properties, as well as the Eckart-Young theorem in \eqr{eq:eckartyoung}, not only make the SVD very mathematically appealing but have also aided in the construction of efficient, accurate and numerically robust algorithms for their computation~\cite{Golub1996}. The $L_2$-norm minimizing rank-$\chi$ approximation $\tilde{\Theta}$ is formed by truncating the inner dimension in the SVD from $X$ to $\chi$.

\subsection{Optimality of factorizations within cTEBD}
\label{sec:optimality}
For the calculation of arbitrary expectation values, we would like to minimize the distance between $\ket{P}$ and its cTEBD approximation $\ket{Q}$. Within cTEBD it is the local $\Theta$ tensor that is factorized, and discussions in the previous section explain how the algorithms will attempt to solve for the optimality of the approximation $\tilde{\Theta}$ to the $\Theta$ tensor. Here, we examine the optimality of this factorization for the distance between $\ket{P}$ and $\ket{Q}$.

A unique property of the $L_2$-norm is that it is unitarily invariant and so we are free to evaluate it in any orthonormal basis. For this reason the SVD truncation $\tilde{\Theta}$ of $\Theta$, which minimizes the corresponding local error $\| \ket{\Theta} - \ket{\tilde{\Theta}} \|_2$, also minimizes the global error $\| \ket{P} - \ket{Q}\|_2$ between the corresponding vectors, so long as the indices of $\tilde{\Theta}$ and $\Theta$ correspond to an orthonormal basis. In \apr{sec:mps_construct} we discuss the orthonormality constraints that the tensors $\{ A^{[\ell]} \}$ need to obey for this to be the case. In the context of classical systems, where inherently non-unitary local stochastic operators are applied~\footnote{A similar situation occurs for imaginary time-evolution of quantum systems and the evolution of mixed states according to a quantum master equation \cite{VidalPrivate}.}, maintaining such orthonormal properties of the tensors $\{ A^{[\ell]} \}$ is by no means guaranteed. Yet it turns out that by applying the two-site operators in a sweeping sequence across the system the relevant bases for each pair are always orthonormal and the SVD truncation is optimal for each operation~\cite{VidalPrivate}. This is described in detail in \fir{fig:orthonormal}. The second-order Suzuki-Trotter expansion introduced earlier is based precisely on such a sweep and is therefore an ideal choice for implementing generic non-unitary evolution on an MPS if the $L_2$-distance is the cost function to be minimized.
\begin{figure}[tb]
\includegraphics[width=8.6cm]{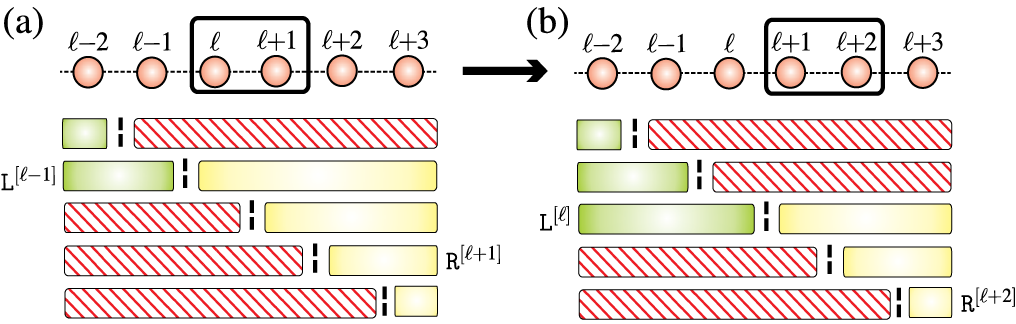}
\caption{(Color online) Following~\cite{VidalPrivate} we illustrate the orthonormality structure of an MPS. Here the boxes correspond to the sets of basis states for the left and right subsystems for different bipartitions (see \apr{sec:mps_construct} for details). If these states form an orthonormal basis then the box is shaded, and correspondingly it is hashed if they do not. (a)~This is the situation obtained by following the procedure described in \fir{fig:trotter_twosite} up until $S_{\ell -1}$ has just been applied and $S_{\ell}$ is next to be applied. All splittings to the left of the one between sites $(\ell -1, \ell)$ have left states which are orthonormal, while all those to the right have their right states orthonormal. The orthogonality centre is said to be located after site $\ell -1$. If the two-site stochastic operator $S_\ell$ is applied to sites $(\ell,\ell+1)$ then we can be assured that all the indices of the $\Theta$ tensor are orthonormal since subsystems ${\tt L}^{[\ell-1]}$ and ${\tt R}^{[\ell+1]}$ have orthonormal bases. (b) This is the situation after the application of $S_{\ell}$. While the action of this non-unitary operator destroys the orthonormality of ${\tt R}^{[\ell-1]}$, the unitarity of the SVD decomposition establishes it in ${\tt L}^{[\ell]}$. This results in the orthogonality centre moving one site to the right. If we applied another two-site operator to sites $(\ell+1,\ell+2)$ then we are again assured of the orthonormality of the tensor indices since subsystems ${\tt L}^{[\ell]}$ and ${\tt R}^{[\ell+2]}$ have orthonormal bases in (b). Finally, analogous changes in the orthonormality structure occur for two-site operations moving to the left.}\label{fig:orthonormal}
\end{figure}  

For cTEBD based on sMPS and NMFs, orthogonality is less useful since in general it cannot be imposed on an sMPS due to its non-negativity, and moreover the $L_1$-norm minimized by the NMF is not unitarily invariant. Despite the lack of optimality the $L_1$-norm error of an NMF of the local $\Theta$ tensor $\tilde{\Theta}$ can be shown to upper-bound the full $L_1$-norm error of the vectors. To see this we use the fact that the full $L_1$-norm between the exact vector $\ket{P}$ and the approximate one $\ket{Q}$ simplifies in matrix product form because the sMPS of the two states differ only for the two sites $(\ell,\ell+1)$ on which an operator is applied,
\begin{eqnarray}
&& \|\ket{P}-\ket{Q}\|_1 \nonumber \\
&&= \sum_{\bf i}\left|A^{[1]i_1}\cdots A^{[\ell-1]i_{\ell-1}}(\Theta^{i_\ell i_{\ell+1}} - \tilde{\Theta}^{i_\ell i_{\ell+1}})A^{[\ell+2]i_{\ell+2}}\cdots A^{[N]i_N}\right|, \nonumber \\
&&\leq \sum_{\bf i}A^{[1]i_1}\cdots A^{[\ell-1]i_{\ell-1}}\left|\Theta^{i_\ell i_{\ell+1}} - \tilde{\Theta}^{i_\ell i_{\ell+1}}\right|A^{[\ell+2]i_{\ell+2}}\cdots A^{[N]i_N}, \nonumber \\
&&=  C^{[1]}\cdots C^{[\ell-1]}\left\{\sum_{i_\ell i_{\ell+1}}\left|\Theta^{i_\ell i_{\ell+1}} - \tilde{\Theta}^{i_\ell i_{\ell+1}}\right|\right\}C^{[\ell+2]}\cdots C^{[N]}, \nonumber \\
&&\leq \sum_{\mu_{\ell-1}\mu_{\ell+1}}\sum_{i_\ell i_{\ell+1}}\left|\Theta^{i_\ell i_{\ell+1}}_{\mu_{\ell-1}\mu_{\ell+1}} -\tilde{\Theta}^{i_\ell i_{\ell+1}}_{\mu_{\ell-1}\mu_{\ell+1}}\right|. \nonumber
\end{eqnarray}  
Here we have used the non-negativity of tensors $\{ A^{[\ell]} \}$ along with the fact that the left $C^{[1]}\cdots C^{[\ell-1]}$  and right $C^{[\ell+2]}\cdots C^{[N]}$ products, where $C^{[\ell]} = \sum_{i_\ell}A^{[\ell]i_\ell}$, form a row and column vector, respectively, composed of non-negative elements $\leq 1$. The last line is then the $L_1$-norm of the local tensor approximation minimized by the NMF, reshaped as a vector.

\section{The TASEP}
\begin{figure*}[tpb]
\includegraphics[width=17.8cm]{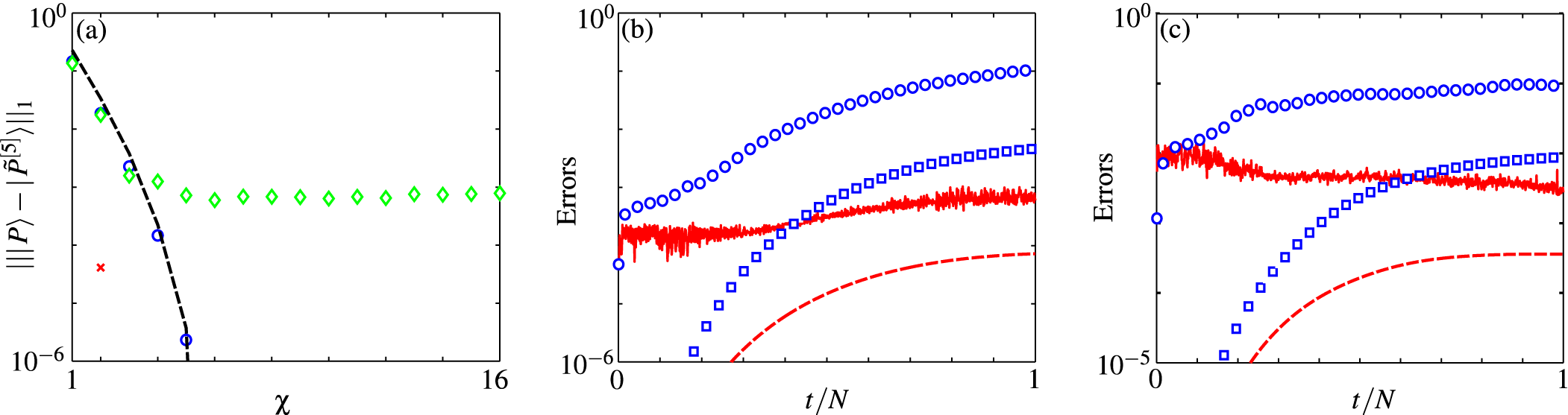}
\caption{(Color online) \label{fig:aseptest}(a) The accuracy of approximations to the stationary state $\ket{P}$ of the 10-site TASEP with $\alpha=\beta = 1$. The $L_1$-norm error is plotted against $\chi$ for the SVD~($\circ$) and $L_1$-norm minimizing NMF ($\diamondsuit$) performed at the central bipartition. For each SVD the $L_1$-norm error due to the sum of all negative elements is shown ($\times$) along with the upper-bound to the $L_1$-norm error (dashed line) derived from the exact $L_2$-norm error. (b) Errors during the evolution of the 10-site TASEP with $\alpha=0.25$ and $\beta = 0.75$. The NMF and SVD algorithms are compared for $\chi=3$ and $\delta t  = 10^{-2}$ where for each NMF four random restarts have been performed. Initially, the system was in the state describing a uniform distribution over all configurations. With decimation occurring at the central bipartition only, we plot the upper-bound to the $L_1$-norm error incurred during each time-step, discussed in \apr{sec:trotter_trunc_errors}, using the SVD (smooth dashed line) and NMF (jagged solid line), and the actual $L_1$-norm error for the SVD ($\Box$) and NMF ($\circ$) at time $t$. (c) The same as (b) but with decimation occurring at all contiguous bipartitions.}
\end{figure*}
\label{sec:ASEP}
Having introduced cTEBD, the rest of this paper deals with analyzing the performance of cTEBD on a test system, the TASEP, which we describe in \fir{fig:ASEP_model}. For a detailed review of the TASEP we refer the reader to~\cite{Golinelli2006,Derrida1998}, but for our purpose it is sufficient to note that the steady state of the system has been solved analytically~\cite{Derrida1993}, while numerical simulations are needed to calculate arbitrary expectation values away from steady state. Like many of the particle hopping systems, the TASEP is modeled as a chain of $N$ lattice sites, each of which can be in one of $d$ configurations, and thus can be described by the formalism introduced in \secr{sec:classical_mps}. For the TASEP $d=2$ and the local configurations $i_\ell=0,1$ correspond to empty and occupied sites respectively. As for a large class of stochastic systems with configurations connected by Poisson processes, the probabilities of being in each complete configuration $\{ P_{\bf i} (t) \}$ evolve according to a master equation of the form
\begin{equation}\label{eq:master_eqn}
\frac{\partial P_{\mathbf{i}} (t)}{\partial t} = \sum_{\mathbf{i}' \neq \mathbf{i}}
\big( 
P_{\mathbf{i}'} (t)
R_{\mathbf{i}' \rightarrow \mathbf{i}}  -P_{\mathbf{i}}(t)
R_{\mathbf{i} \rightarrow \mathbf{i}'} 
\big) ,
\end{equation}
where $R_{\mathbf{i} \rightarrow \mathbf{i}'} $ is the Poisson rate governing the transition from $\mathbf{i}$ to $\mathbf{i}'$~\cite{Blythe2007}. The first term on the right hand side gives the rate of transitions into configuration $\mathbf{i}$ and the second gives the rate of leaving it.

We may rewrite such a master equation \eqr{eq:master_eqn} as a stochastic Schr\"{o}dinger equation~\cite{Glauber1963,Alcaraz1993} of the form
\begin{equation}\label{eq:schrodinger_eqn}
\frac{\partial}{\partial t} \ket{P\left(t\right)}= H\ket{P\left(t\right)}, \nonumber
\end{equation}
where the stochastic Hamiltonian $H$ is defined by
\begin{subequations}\label{eq:hamiltonian_def}
\begin{align}\label{eq:hamiltonian_def1}
  \bra{\mathbf{i}}H\ket{\mathbf{i}'} &= R_{\mathbf{i}' \rightarrow \mathbf{i}}  \; \; \; \mathrm{for} \; \; \; \mathbf{i}\neq \mathbf{i}' ,\\\label{eq:hamiltonian_def2}
  \bra{\mathbf{i}}H\ket{\mathbf{i}} &= - \sum_{\mathbf{i}' \neq \mathbf{i}} R_{\mathbf{i} \rightarrow \mathbf{i}'}  .
\end{align}
\end{subequations}
The probability vector describing stochastic systems therefore undergoes non-unitary evolution via a Schr\"{o}dinger equation in imaginary time with a non-Hermitian stochastic Hamiltonian. From Eqs. (\ref{eq:hamiltonian_def}) a stochastic Hamiltonian has non-negative off-diagonal elements, a consequence of the non-negativity of transition rates, and non-positive diagonal elements to ensure the conservation of probability 
\begin{equation}\label{eq:conservation_prob}
\sum_{\mathbf{i}'} \bra{\mathbf{i}'} H \ket{\mathbf{i}} = 0. \nonumber
\end{equation}
It has been shown in~\cite{Keizer1972} that there always exists a stationary state, an eigenvector of $H$ corresponding to a zero eigenvalue, and this vector is unique if $H$ is strongly connected, i.e. all configurations are accessible, which is true of the TASEP. This implies ergodicity; after long times it will arrive at a distribution independent of its initial state. The Hamiltonian for the TASEP can be written as
\begin{equation}\label{eq:total_hamiltonian}
H = h_1 + h_N + \sum_{\ell=1}^{N-1} h_{\ell,\ell+1},
\end{equation}
where the single-site terms $h_1$ and $h_N$ describe the input and output of particles and the two-site nearest-neighbor terms $\{ h_{\ell,\ell+1} \}$ describe the hopping. The Hamiltonian terms can be calculated using Eqs. (\ref{eq:hamiltonian_def}) and are given in~\cite{Rajewsky2004}. So evolution is generated by a stochastic operator of the form considered in \secr{sec:trotter}. 

For simplicity we take time $t$ to be in units of the inverse of the hopping rate $\gamma^{-1}$ and set $\gamma = 1$. Two observables that we will be particularly interested in are the density at site $\ell$, $\rho_\ell$, which is equal to the local configuration $i_\ell$, and the current between site $\ell$ and $\ell + 1$, $J_\ell$, equal to $1$ if $i_\ell = 1,i_{\ell + 1} = 0$, and $0$ otherwise.

\section{Small system analysis}\label{sec:small_sys}
\subsection{Comparison of the NMF and SVD for the TASEP}
\begin{figure*}[tpb]
\includegraphics[width=17.8cm]{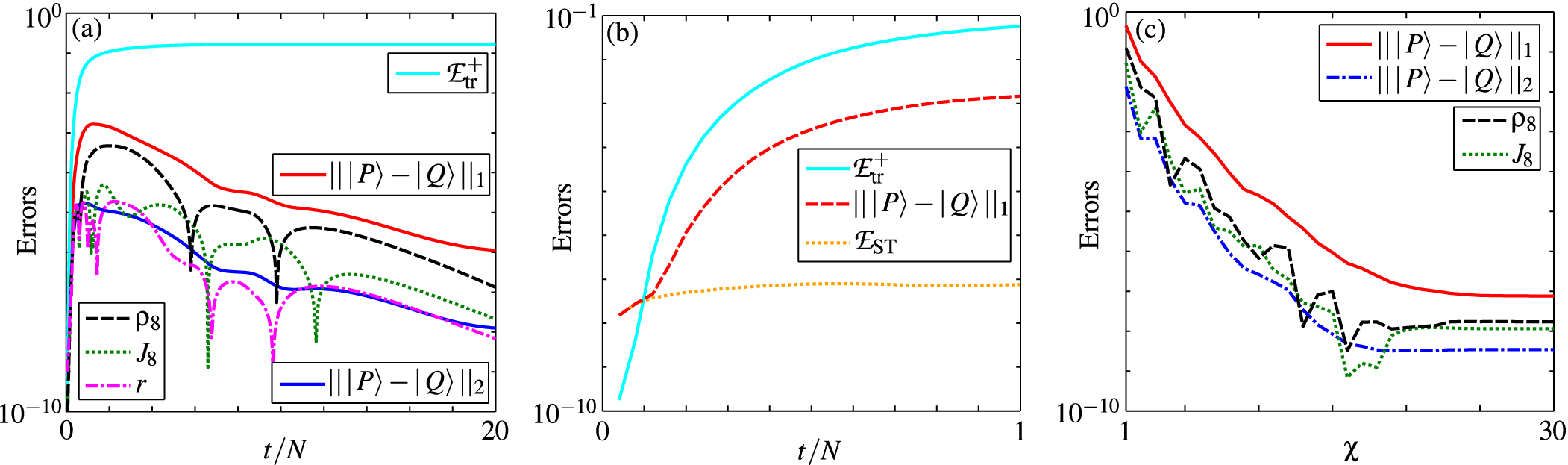}
\caption{(Color online) \label{fig:obserror}The errors in simulating the 15-site TASEP with $\alpha=0.3$ and $\beta=0.6$, using cTEBD with $\delta t = 10^{-3}$, as compared to an exact simulation. (a) Using $\chi=5$ we plot the distances between the approximate and exact probability vectors, $|| \ket{P}- \ket{Q} ||_1$ and $|| \ket{P}- \ket{Q} ||_2$ along with the errors $\left| \langle O \rangle^P - \langle O \rangle^Q \right|$ for each observable $O$. The upper-bound $\mathcal{E}_{\textrm{tr}}^{+}$ is also plotted and of the solid lines this takes the largest values, with the $L_1$ and then $L_2$-norm errors below. (b) A zoom in on the small time part of (a) along with the $L_1$-norm error due to the Trotter expansion $\mathcal{E}_{\textrm{ST}}$ calculated using a $\chi = 100$ simulation. (c) The behavior of some of these distance measures with increasing $\chi$ at $t = N$.}
\end{figure*}
A physically relevant comparison of the NMF and SVD can be made by considering the probability distribution of the TASEP stationary state. The $L_1$-norm errors for the SVD and NMF approximations to this probability distribution bipartitioned at centre of the system are presented in \fir{fig:aseptest}(a) for $\alpha = \beta = 1$ and $N=10$. This shows that the SVD is exact up to machine precision once $\chi > 5$ indicating that the stationary state has weak enough correlations to permit a considerable amount of compression. The NMF fails to identify an accurate solution as $\chi$ is increased, due to issues with local minima, and gives an error which levels off above $10^{-3}$. Note also that negativity of the approximation found from the SVD only occurs for $\chi = 2$ and is several orders of magnitude smaller than the overall $L_1$-norm error. Finally, the upper-bound to the $L_1$-norm error for the SVD truncation, computed from the $L_2$-norm error, is also plotted and seen to provide a reasonably tight upper-bound.

To move beyond a single matrix factorization we have also tested how the NMF and SVD behave when used repeatedly to simulate time-evolution. We applied cTEBD to the TASEP model, in which the two sets of 5 sites on either side of the central bipartition of a 10-site system were merged so only this single bipartition was considered. A calculation of the time-evolution was performed up to a time $t = N$, using both factorizations with $\chi=3$. At this level of truncation the SVD performed roughly the same as the NMF in the stationary state, as seen in \fir{fig:aseptest}(a). The results for this time-evolution are shown in \fir{fig:aseptest}(b). Initially, the system is in a $\chi=1$ state but the NMF fails to find accurate decompositions at small times, while the SVD succeeds. At later times the difference in errors at each time-step reduce to an order of magnitude. To examine the optimality issue explained in \secr{sec:optimality} we have repeated the calculation, this time using a full 10-site matrix product decimation. We found that while both NMF and SVD errors are slightly worse when decimation occurs at each bipartition, this is not more serious for the NMF than the SVD algorithm, as demonstrated in \fir{fig:aseptest}(c). Rather than optimality, the problem faced using current NMF algorithms is the plateau in accuracy seen for increasing $\chi$. This not only prevents any improvements, in contrast to the SVD, but causes unnecessary accumulation of error. Also to support the use of the SVD, it is found that even with $\chi = 3$ the evolution never produces any negative probabilities over this time. These results strongly suggest that the potential for negativity when using the SVD within the cTEBD framework does not present a serious obstacle. For this reason we shall, from this point on, focus on SVD-based decimation.

\subsection{Behavior of cTEBD errors}
\label{sec:ctebderrors}
We now examine in detail the behavior of the different types of errors of SVD-based cTEBD for small systems, in preparation for studying the behavior of the same errors for larger systems in the next section. As a demonstration of the relationship between different error measures we show in \fir{fig:obserror}(a) the $L_1$ and $L_2$-norm errors of a simulation of the initially empty $15$-site TASEP over a time $t=20N$. Also in the figure we have plotted the expectation value errors of three observables: an observable $r$ whose value at each configuration was randomly generated from a uniform distribution between $0$ and $1$; the density at the middle site $\rho_8$; and the current leaving the middle site $J_8$. The random observable scales with the $L_2$-norm error, because it is uncorrelated with the MPS approximation method, as expected from the discussion of observable error scaling in \apr{sec:obs_errors}. The current scales slightly above this, and the density lies even closer to the $L_1$-norm error, indicating that there is some correlation between cTEBD's errors and these two observables. The $L_1$-distance lies close to the maximum factor of $2^{N/2}$ above the $L_2$-distance and, as must be the case, all errors are bounded by the $L_1$-distance.

We may isolate the Trotter error by using a $\chi$ large enough such that the truncation error is negligible. Also, it is always possible to efficiently calculate $\mathcal{E}_{\textrm{tr}}^{+}$, introduced in \apr{sec:trotter_trunc_errors}, which upper-bounds the cumulative $L_1$-norm error due to truncation. The interplay between truncation and Trotter errors can be clearly seen in \fir{fig:obserror}(b), which shows that the $L_1$-distance follows the Trotter error, until a time at which truncation errors become more significant, as indicated by the increasing bound. $\mathcal{E}_{\textrm{tr}}^{+}$ is an overestimate of $\mathcal{E}_{\textrm{tr}}$ and after times of a few $N$ it plateaus, as can be seen in \fir{fig:obserror}(a). As it is cumulative, the final value of $\mathcal{E}_{\textrm{tr}}^{+}$ can be used to bound the $L_1$-distance between the cTEBD and exact probability distributions, and so all errors, up to this time.
\begin{figure*}[tbh]
\includegraphics[width=17.8cm]{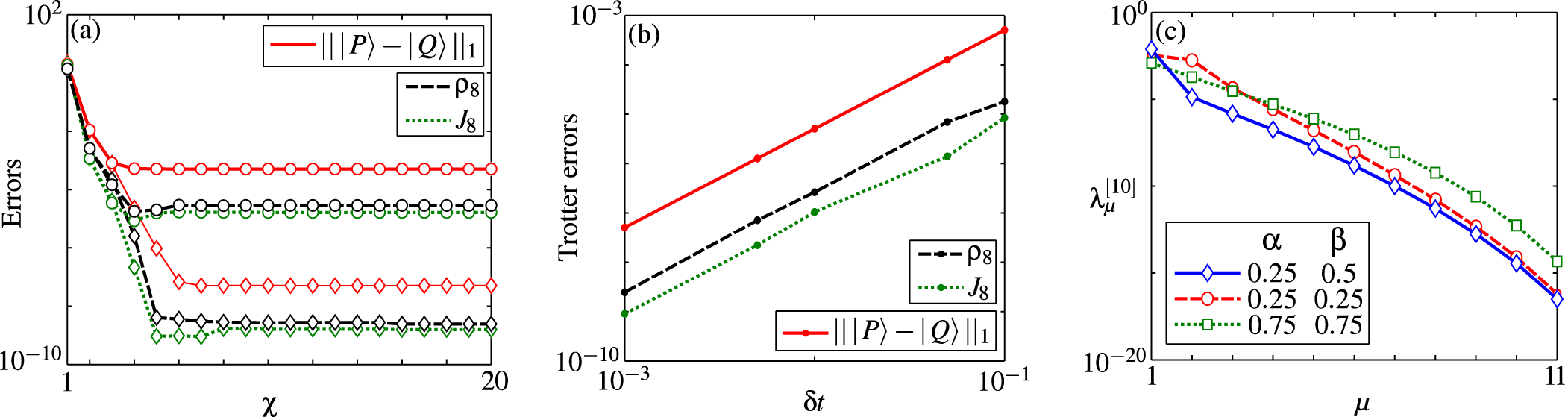}
\caption{(Color online) \label{fig:sserror}The steady state of the TASEP. (a) For the 15-site TASEP with $\alpha=0.3$ and $\beta=0.6$ we have plotted the convergence of error measures to the Trotter error as $\chi$ is increased. This is shown for $\delta t = 10^{-1}$ ($\circ$) and $\delta t = 10^{-3}$ ($\diamondsuit$). For both observables $O$ we have plotted $\left| \langle O \rangle^P - \langle O \rangle^Q \right|$. (b) For the same system as (a) this shows the Trotter errors, calculated from the $\chi=20$ simulations, plotted against $\delta t$ on a log-log axis. To guide the eye, we have connected the points with lines. They show a $\delta t^2$ dependence. (c) The steady state singular value spectra of the 20-site TASEP, corresponding to bipartite splittings through the middle of the system. Above $\mu=11$ the singular values are integer zero.}
\end{figure*}

Another consequence of $\mathcal{E}_{\textrm{tr}}^{+}$ being cumulative is that it has less relevance for much longer times, when the system approaches steady state, especially since the errors do not increase monotonically in time. Consider the evolution of the TASEP shown in \fir{fig:obserror}(a). Initially the system is in a specific configuration, in this case the empty lattice, which can be exactly represented by an MPS with $\chi=1$ and so there is zero error. In time, correlations are introduced into the system and the transient errors grow to a peak located at $t \approx N$, before decreasing as the steady state is approached, suggesting that the steady state is particularly compressible. This shows that even though a low $\chi$ may not be sufficient to describe the transient behavior of the system accurately, it could still give accurate results in the steady state. This is because the TASEP is ergodic~\cite{Keizer1972} and the Trotter approximation to the evolution operator in \eqr{eq:Trotter} appears to preserve this property. So through whichever states the low $\chi$ approximation to the transient behavior takes the system, whether this be erroneous or not, the approximation to the steady state will be the same.

For small systems we can use cTEBD to calculate any observable of the TASEP precisely, because by increasing $\chi$, while still severely compressing the system, we can restrict the $L_1$-norm error to small values. This is shown, for a time $t=N$ near the peak in errors, in \fir{fig:obserror}(c). As $\chi$ is increased, truncation errors shrink until the errors are dominated by the Trotter error. Even though both the $L_1$ and $L_2$-norm errors seem to decrease monotonically with $\chi$, from \fir{fig:obserror}(c) it is clear this is not necessarily the case for observable errors. This non-monotonic behavior arises because the correlation between errors in the probability vector and an observable's values may vary with $\chi$.

The behavior of the errors in time, shown in \fir{fig:obserror}(a), suggests that the steady state is more compressible than the transient states. To confirm this we evolved the TASEP up to $t = 100N$ by which time the system was effectively in the steady state. For this state, we show in \fir{fig:sserror}(a) that errors can be driven down to the Trotter error for $\chi$ even as small as $4$ when using a time-step $\delta t=10^{-1}$. To obtain more accurate results, $\delta t$ can be decreased, and the results presented in \fir{fig:sserror}(b) confirm that the scaling of $\mathcal{E}_{\textrm{ST}}$ is with $\delta t^2$. The success of the cTEBD algorithm in describing the TASEP steady states accurately with an MPS of small $\chi$ stems from the rapid decay of the singular value spectra of the bipartite splittings of the probability distribution. In~\cite{Temme2010} an exact sMPS for every $N$-site TASEP steady state was obtained with a dimension $\chi = N + 1$. From our discussion in \secr{sec:classical_mps} we know an exact SVD-MPS must exist of a dimension less than or equal to that of an exact sMPS, and this can be constructed explicitly by repeatedly performing SVDs on the matrices of the sMPS given in~\cite{Temme2010}. By doing this we have plotted, in \fir{fig:sserror}(c), the spectra of singular values for the 20-site TASEP for a few values of $\alpha$ and $\beta$. The spectra decay super-exponentially before becoming integer zero after $\mu = N/2 +1$. Numerical tests suggest that this is the matrix dimension needed for an exact MPS. How these properties extend to larger systems is investigated in the next section.
\begin{figure*}[tbh]
\includegraphics[width=17.8cm]{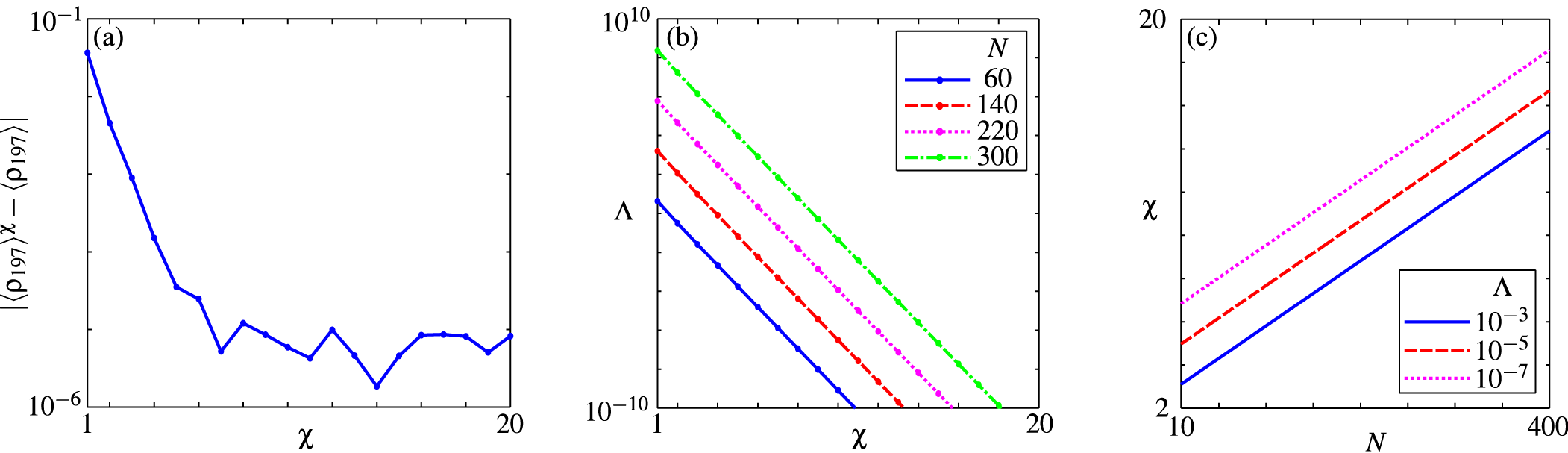}
\caption{(Color online) \label{fig:bigsysa}(a) The convergence of the cTEBD density expectation value $\langle \rho_{197} \rangle^\chi$ to the analytical result $\langle \rho_{197} \rangle$ for $N=200$. The system is in the steady state for $\alpha=0.349$, $\beta=0.537$, having used $\delta t = 5 \times 10^{-3}$. (b) $\Lambda$ plotted for different $\chi$ and $N$, from the $\alpha = 0.3$, $\beta = 0.6$ steady state, calculated using the exact results from \cite{Temme2010}. (c) For the same system as in (b) this shows the MPS dimension $\chi$ needed to ensure an SVD-MPS exists with $L_1$-distance less than $\Lambda$, as a function of $N$.}
\end{figure*}
\section{Scalability}
\label{sec:scalability}
\begin{figure*}[tbh]
\includegraphics[width=17.8cm]{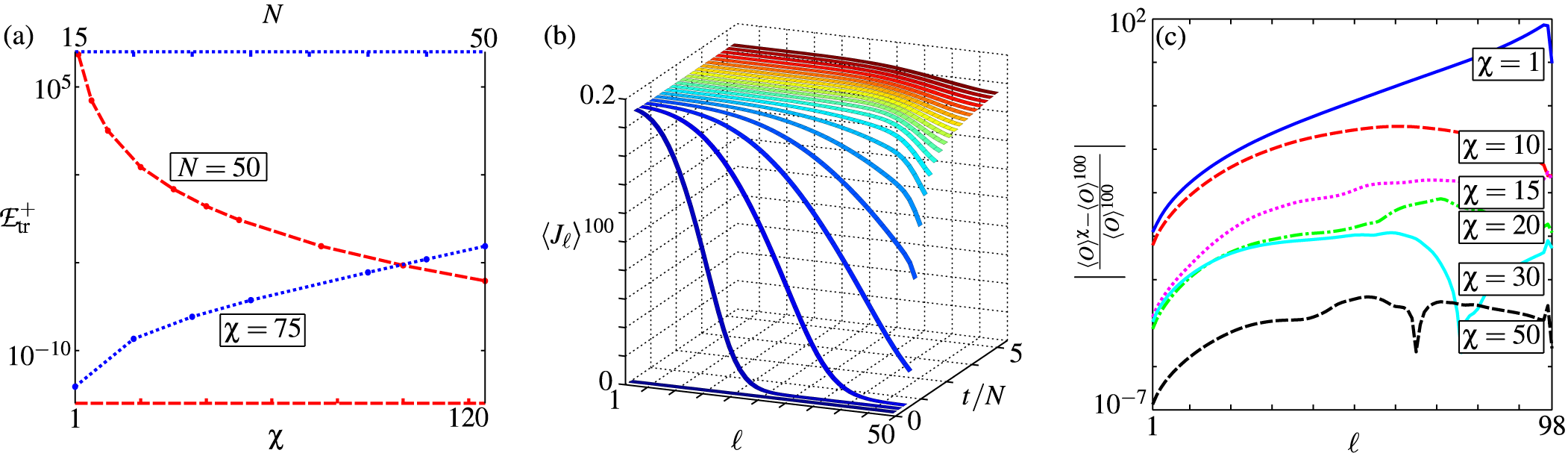}
\caption{(Color online) \label{fig:bigsysb}(a) The dependence of $\mathcal{E}_{\textrm{tr}}^{+}$ on $N$ and $\chi$, corresponding to the upper and lower x-axes respectively. We used $\alpha=0.3$ and $\beta=0.6$, taking the initial state to be the empty lattice, and setting $\delta t = 5 \times 10^{-3}$. Each simulation was ran up to $t = 10 N$ by which time the value of $\mathcal{E}_{\textrm{tr}}^{+}$ had plateaued. (b)~The evolution of the current expectation values, calculated with $\chi=100$, for $N=50$, $\alpha=0.25$ and $\beta=0.25$. The errors are bounded by $\mathcal{E}_{\textrm{tr}}^{+}$ which plateaus at $6 \times 10^{-6}$. (c) The errors between the expectation values of the products of the densities of three sequential sites $\{ O = \rho_\ell \rho_{\ell+1} \rho_{\ell + 2} \}$ and the $\chi=100$ result, showing a convergence to within $10^{-5}$ as $\chi$ is increased to $50$. We used $\alpha=0.25$ and $\beta=0.25$ and calculated expectation values at $t=N$. The time-step used was $\delta t = 5 \times 10^{-3}$, and at $t=0$ the lattice was empty.}
\end{figure*}

The calculation of the exact $L_1$-distance is intractable for large systems. To overcome this we use three approaches to determine the accuracy of calculated expectation values: Firstly, in the steady state we compare expectation values calculated using cTEBD directly with exact analytical results; Secondly, away from steady state, we use $\mathcal{E}_{\textrm{tr}}^{+}$ to upper-bound the $L_1$-distance; Finally, we show for specific observables that as $\chi$ is increased each expectation value converges. 

With the first approach in mind, we simulated the TASEP on a $200$-site lattice for a total time of $100N$, enough to reach the steady state, which, from our analysis of small systems, is when we expect the most accurate results to be obtained. In \fir{fig:bigsysa}(a) we have plotted a comparison between cTEBD and analytical expectation values for the density at a site close to the exiting boundary of the system. Even for such a large system the error is Trotter limited for a $\chi$ of $7$. This observable is actually unusual. For the vast majority of expectation values considered, e.g. current, density and arbitrary two-site correlations, the Trotter error was reached for $\chi=1$, signifying the extreme compressibility of these states, also responsible for the success of mean field theory on these systems. It is also worth noting that the non-monotonic behavior of errors with $\chi$, observed for smaller systems, is present for larger systems.

To understand this compressibility of the steady states it is instructive to calculate the bound on the $L_1$-norm $\Lambda$, introduced in \eqr{eq:Lambda}, using the method for calculating singular spectra from the exact stationary sMPS~\cite{Temme2010}, discussed in \secr{sec:ctebderrors}. This gives us an upper-bound to the $L_1$-distance that may be achieved by an SVD-MPS constructed in the way described in \apr{sec:mps_construct}. Although this construction is not exactly what is performed by cTEBD, the accuracy of both depend on the same properties. We have calculated $\Lambda$ analytically and plotted the results in \ref{fig:bigsysa}(b) for several system sizes. Except for some super-exponential decay when $\chi \approx N/2 +1$ it exhibits a nearly perfect exponential decay in $\chi$ and exponential increase in $N$ of the form $\Lambda = a^{N}b^{-\chi}$, with $a=1.0770$ and $b=11.7797$ respectively. Hence the MPS dimension needed to describe the steady state with errors less than $\Lambda$ is given by $\chi = N \; \mathrm{log}_b (a) - \mathrm{log}_b{\Lambda}$, as plotted in \fir{fig:bigsysa}(c). We find that accurate MPS approximations to the steady states of large systems can be obtained using a $\chi$ of order $10$, due to the rapid decay of the singular value spectra of the steady states, in line with the cTEBD results we have obtained. 

A potential limitation to the scalability of the cTEBD algorithm using the SVD is the exponential scaling with $N$ of the prefactor bounding the $L_1$-norm via the $L_2$-norm, which appears in \eqr{eq:Lambda}. This indicates that in order to maintain a given bound on the $L_1$-norm the simulation must be performed with exponentially increasing $L_2$-norm accuracy with $N$. The extremely rapid decay of the singular spectrum for TASEP stationary states indicates that an $L_1$-norm bound can be maintained by only a very moderate increase in $\chi$. However, we note that in order to exploit this property it may be necessary to use higher precision arithmetic than the standard 64-bit double precision. Indeed, for even moderately large systems, above 50 sites or so, all but the first few singular values have a ratio to the largest that is below double machine precision. TEBD will not be able to reproduce these values. This is a similar problem to what was experienced in~\cite{Carlon1999} for the Reaction-Diffusion model, but in our case numerical instabilities do not arise from this imprecision.

When away from the steady state, we cannot calculate the exact spectrum or expectation values. However, we can calculate $\mathcal{E}_{\textrm{tr}}^{+}$ and we have done this for a range of $N$ and $\chi$ to produce \fir{fig:bigsysb}(a). The results show that to ensure a small error, a larger $\chi$ is needed for larger $N$. We can also deduce, for instance, that the error in calculating the expectation value of any observable for $N=50$ using $\chi =75$ will not exceed $10^{-4}$. In light of this we have plotted in \fir{fig:bigsysb}(b) the time-evolved current profile of an initially empty $50$-site TASEP for a time $t \approx 5N$ using a $\chi$ of $100$. The system fills up as particles injected into the first site hop across the system, and steady state values are almost reached by $t=5N$. From $\mathcal{E}_{\textrm{tr}}^{+}$ we can be sure that all values are correct to a factor of $10^{-5}$, despite evolving through the transient behavior where the largest errors are expected to appear. Note that in this argument we have ignored the Trotter errors, which we have found to much smaller than the truncation errors, and are easily controlled by reducing $\delta t$.

Importantly, $\mathcal{E}_{\textrm{tr}}^{+}$ is an upper-bound, and so it is sufficient, but not necessary for it to take small values. As we saw in \fir{fig:obserror}(a) this upper-bound is often a gross overestimation of the actual $L_1$-norm error as it is the cumulative sum of upper-bounds to the $L_1$-norm error in each two-site operation, as detailed in \apr{sec:trotter_trunc_errors}. Motivated by this, we have used cTEBD to simulate the non-equilibrium dynamics of larger systems, for which $\mathcal{E}_{\textrm{tr}}^{+}$ alone could not guarantee accurate results. We considered how quickly expectation values converge with $\chi$, and calculated this convergence for systems at $t=N$, which in the investigation of small systems was found to be a time close to that with the largest error. As shown in \fir{fig:bigsysb}(c), a numerically calculated expectation value for $N=100$ converges such that the $\chi=50$ and $100$ results differ only by a fraction of $10^{-5}$. So convergence occurs such that only a moderate $\chi$ is needed to accurately calculate the expectation value of this observable, even for transient states. This observable for which the results are plotted could, for example, be used to indicate the presence of a traffic jam with the TASEP used to model a single carriageway. Similar results were obtained for a range of observables, indicating that the cTEBD algorithm is able to accurately simulate the transient dynamics of the TASEP for these system sizes.

\section{Conclusions}
\label{sec:conclusions}
In this paper we have investigated in detail the application of the TEBD algorithm to classical stochastic systems. A very reasonable approach to describe probability distributions is to use an sMPS~\cite{Temme2010} which is manifestly non-negative. We have shown that cTEBD can produce approximations to the dynamics of stochastic systems within the sMPS class by modifying only the factorization method, namely switching it to NMF. Unfortunately, current NMF algorithms find it difficult to locate accurate low rank solutions, due to the non-convexity of the factorization. However, issues of non-optimality due to the non-orthogonality of the MPS do not seem to cause problems, so if future algorithms overcome the former problem NMF-based cTEBD could become a practical option. The main approach explored here was to relax the constraint to be explicitly non-negative by using instead a general real MPS and applying the SVD method for the evolution in an essentially identical fashion to quantum problems. We showed that potential issues involving the proliferation of negativity and that the $L_2$ rather than $L_1$-distance is minimized turn out not to present serious obstacles to the SVD-based algorithm for the system sizes considered.

With the TASEP as our test system, we demonstrated the accuracy and applicability of cTEBD for non-equilibrium systems. We focused our investigation on the behavior of errors with truncation $\chi$ and system size $N$. An interesting feature was the lack of a monotonic dependence of expectation value errors on $\chi$, though in all cases considered, with system sizes up to the low hundreds, results converged. For the steady state, comparisons with analytically calculated expectation values were excellent and we obtained negligible errors for $\chi$ less than $10$. Away from steady state we introduced an upper-bound to the $L_1$-distance, and thus all errors, which could be efficiently calculated as part of the algorithm when simulating any system, not just the TASEP. This was used to ensure that some of our $50$-site simulations had negligible errors. Even for large systems for which this bound could not guarantee small errors, we found that expectation values converged for a large range of observables.

This work suggests that cTEBD is a viable candidate for studying the time-evolution of a large class of non-equilibrium stochastic systems whose dynamics are governed by a stochastic Hamiltonian consisting of at most two-site nearest-neighbor terms. Since cTEBD is easily adjusted to include many rates with time and space dependence as well as being extendable to network geometries with a bounded tree-width~\cite{Shi2006,Markov2009}, it could prove to be a powerful simulation option. The accuracy of cTEBD shown here also indicates that there is great scope to adapt other more sophisticated MPS/tensor network methods to classical stochastic systems and extend simulations to higher dimensions. These include the use of matrix product operators to express the evolution for longer ranged processes~\cite{Huebener2010,Frowis2010,Banuls2009,Stoudenmire2010} and renormalization inspired variational techniques, which have already been used with great success for both stationary and dynamical calculations of quantum systems~\cite{Verstraete2009,Cirac2009}. Future work will investigate the performance of cTEBD and its extensions to a wider range of systems and geometries.

\begin{acknowledgments}
SRC and DJ thank the National Research Foundation and the Ministry of Education of Singapore for support. DJ acknowledges support from the ESF program EuroQUAM (EPSRC grant EP/E041612/1), the EPSRC (UK) through the QIP IRC (GR/S82176/01), and the European Commission under the Marie Curie programme through QIPEST. TJ thanks Vlatko Vedral for useful discussions.
\end{acknowledgments}

\appendix

\section{Construction of an MPS via repeated SVDs}
\label{sec:mps_construct}
\begin{figure}[b]
\begin{center}
\includegraphics[width=7cm]{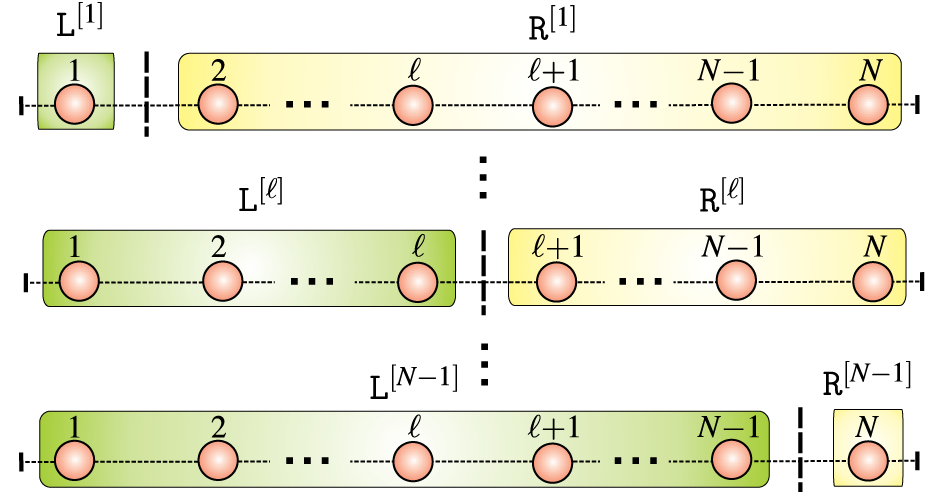}
\caption{(Color online) The sequence of $N-1$ contiguous partitions (according to
the labeling imposed) of the system in which the Schmidt
decompositions are computed.} \label{fig:schmidt}
\end{center}
\end{figure}
In this section we follow Vidal~\cite{Vidal2003} and show how through the repeated use of SVDs (Schmidt decompositions) an MPS of any vector $\ket{\psi}$ can, in principle, be found. To do this we first compute the Schmidt decomposition of $\ket{\psi}$ for every contiguous bipartition into blocks ${\tt L}^{[\ell]}$ and ${\tt R}^{[\ell]}$, as depicted in \fir{fig:schmidt}. Starting from the left boundary, each left Schmidt state $\kets{{\tt L}^{[\ell]}_{\,\mu_\ell}}$ is iteratively expanded in terms of the local basis of the rightmost site of the block $\ket{i_\ell}$ and the left Schmidt states $\{ \kets{{\tt L}^{[\ell-1]}_{\,\mu_{\ell-1}}} \}$ of the neighboring splitting one site further to the left. This gives the following set of expansions
\begin{eqnarray}
\kets{{\tt L}^{[1]}_{\,\mu_1}} &=& \sum_{i_1=0}^{d-1} A^{[1]i_1}_{\mu_1}\ket{i_1}, \nonumber \\
\kets{{\tt L}^{[2]}_{\,\mu_2}} &=& \sum_{i_2=0}^{d-1}\sum_{\mu_1=1}^{X_1} A^{[2]i_2}_{\mu_1\mu_2}\kets{{\tt L}^{[1]}_{\,\mu_1}}\ket{i_2}, \nonumber \\
&\vdots& \nonumber \\
\kets{{\tt L}^{[\ell]}_{\,\mu_\ell}} &=& \sum_{i_\ell=0}^{d-1}\sum_{\mu_{\ell-1}=1}^{X_{\ell-1}} A^{[\ell]i_\ell}_{\mu_{\ell-1}\mu_\ell}\kets{{\tt L}^{[\ell-1]}_{\,\mu_{\ell-1}}}\ket{i_\ell}, \nonumber \\
&\vdots& \nonumber \\
|{\tt L}^{[N]}_{\,\mu_N=1}\rangle &=& \ket{\psi} =  \sum_{i_N=0}^{d-1} \sum_{\mu_{N-1}=1}^{X_{N-1}} A^{[N]i_N}_{\mu_{N-1}} \kets{{\tt L}^{[N-1]}_{\,\mu_{N-1}}}\ket{i_N}. \label{eq:last_term}
\end{eqnarray}
The expansion coefficients can be seen to define the tensors $\{ A^{[\ell]} \}$ of an MPS representation of $\ket{\psi}$ by taking the last expansion of $\ket{\psi}$ in \eqr{eq:last_term} and inserting all the others into it until no Schmidt states remain
\begin{equation}
\ket{\psi} = \sum_{\bf i}\left\{\sum_{\mu_1=1}^{X_1}\sum_{\mu_2=1}^{X_2}\cdots\sum_{\mu_{N-1}=1}^{X_{N-1}} A^{[1]i_1}_{\mu_1}A^{[2]i_2}_{\mu_1\mu_2}\cdots A^{[N]i_N}_{\mu_{N-1}}\right\}\ket{\bf i}. \nonumber
\end{equation}
The term in the parentheses is identical to the matrix product expansion of the amplitudes $\psi_{\bf i}$ given in \eqr{eq:mps}, but with all the summations written out explicitly. Truncating each summation (equivalent to truncating the Schmidt decomposition or SVD) to $\chi$ terms results in $\ket{\tilde{\psi}}$, an approximate MPS of dimension $\chi$.

While the full product of matrices $\{ A^{[\ell] i_\ell} \}$ generates $\ket{\psi}$, a partial product up to a site $\ell$ instead forms an MPS of a left Schmidt state $ \kets{{\tt L}^{[\ell]}_{\,\mu_\ell}}$ as
\begin{equation}
\kets{{\tt L}^{[\ell]}_{\,\mu_\ell}} = \sum_{{\bf i}} (A^{[1]i_1}A^{[2]i_2}\cdots A^{[\ell] i_\ell})_{\mu_\ell}\ket{{\bf i}}_{\tt L}. \nonumber
\end{equation} 
The orthonormality of the left Schmidt states results in all the tensors $\{ A^{[\ell]} \}$ derived from this construction obeying a left orthonormality property~\cite{Daley2004}
\begin{equation}
\sum_{i_\ell=0}^{d-1} (A^{[\ell]i_\ell})^\dagger A^{[\ell]i_\ell} = \mathbbm{1}_{\chi_{\ell}}, \label{eq:left_orth}
\end{equation}
where $\mathbbm{1}_{\chi_{\ell}}$ is the $\chi_{\ell} \times \chi_{\ell}$ identity matrix. Notice that we could have equally performed this construction starting from the right boundary and using the right Schmidt states $\{ \kets{{\tt R}^{[\ell]}_{\,\mu_\ell}} \}$. This would result in a different set of tensors $\{ A^{[\ell]} \}$ obeying a right orthonormality property
\begin{equation}
\sum_{i_\ell=0}^{d-1} A^{[\ell]i_\ell} (A^{[\ell]i_\ell})^\dagger = \mathbbm{1}_{\chi_{\ell-1}}. \label{eq:right_orth}
\end{equation}
If we were to apply the left procedure up to and including site $\ell$ and the right procedure up to an including site $\ell + 1$ then the resulting MPS would be of the form
\begin{equation}
\ket{\psi} = \sum_{\bf i} A^{[1]i_1}A^{[2]i_2}\cdots A^{[\ell] i_\ell} D^{[\ell]} \, A^{[\ell+1] i_{\ell+1}}\cdots A^{[N]i_N} \ket{\bf i}, \label{eq:mps_twist}
\end{equation}
where $D^{[\ell]}$ is the diagonal matrix of singular values $\{ \lambda^{[\ell]}_{\mu_\ell} \}$. This MPS has an orthogonality centre, or twist in its handedness, located precisely at the bipartition after site $\ell$, since the constraints \eqr{eq:left_orth} and \eqr{eq:right_orth} are obeyed by the tensors to the left and right of $D^{[\ell]}$, respectively. This is crucial since we can readily extract from the MPS in \eqr{eq:mps_twist} an expansion of $\ket{\psi}$ in an orthonormal basis
\begin{equation}
\ket{\psi} = \sum_{\mu_{\ell-1},\mu_{\ell+1}}\sum_{i_\ell,i_{\ell+1}}\psi^{i_\ell i_{\ell+1}}_{\mu_{\ell-1}\mu_{\ell+1}} \kets{{\tt L}^{[\ell-1]}_{\,\mu_{\ell-1}}}\ket{i_\ell}\ket{i_{\ell+1}}\kets{{\tt R}^{[\ell+1]}_{\,\mu_{\ell+1}}}, \nonumber
\end{equation}
by forming the two-site tensor about the orthogonality centre
\begin{equation}
\psi^{i_\ell i_{\ell+1}}_{\mu_{\ell-1}\mu_{\ell+1}} = \sum_{\mu_\ell} A^{[\ell]i_\ell}_{\mu_{\ell-1}\mu_\ell} \lambda^{[\ell]}_{\mu_\ell}A^{[\ell+1]i_{\ell+1}}_{\mu_\ell \mu_{\ell+1}}. \nonumber
\end{equation}
It is only for this pair of sites, or $(\ell-1,\ell)$ and $(\ell+1,\ell+2)$ adjacent to the orthogonality centre, that a tensor of amplitudes for $\ket{\psi}$ can be formed from components of the MPS in \eqr{eq:mps_twist}, such that all indices correspond to orthonormal states. As discussed in \secr{sec:optimality} moving the orthogonality centre, so as to ensure that it is always located adjacent to any pair of sites on which a two-site operator is applied, is essential for the optimality of the subsequent SVD decimation.

\section{Observable errors} \label{sec:obs_errors}
Here, we consider how the errors in calculating an observable using an approximate probability distribution, such as that given by a truncated MPS, should scale with the $L_1$ and $L_2$-distances. Let $\ket{P}$ be the exact probability vector and $\ket{Q}$ be an approximate probability vector. Now consider an observable $O$ taking a value $O_{\bf i}$ for each configuration ${\bf i}$. Since the expectation value is calculated as $\langle O \rangle^P = \sum_{\bf i} O_{\bf i} P_{\bf i}$, the difference between the expected values of the observable, calculated using the two probability vectors is then
\begin{equation}\label{eq:error}
 \langle O \rangle^P - \langle O \rangle^Q =  \sum_{\mathbf{i}} O_{\mathbf{i}} \left( P_{\mathbf{i}} - Q_{\mathbf{i}} \right) , 
\end{equation}
The worst case scenario for this error is that the observable values and the values of the probability error for each configuration are correlated such that each term in the sum is of the same sign. In this case the magnitude of the error can equal its upper-bound
\begin{align}
| \langle O \rangle^P - \langle O \rangle^Q | &\le \sum_{\mathbf{i}} |O_{\mathbf{i}} | \;  | P_{\mathbf{i}}- Q_{\mathbf{i}} |  , \label{eq:l1scaling} \nonumber \\
&\le \max_{\mathbf{i}} \{ O_{\mathbf{i}} \}  || \ket{P} - \ket{Q}  ||_1  . \nonumber
\end{align}
So the error in calculating an expectation value of an observable is always bounded by the $L_1$-norm error. However, for errors to scale as badly as this requires complete correlation between the approximation method and the observable. For many observables we expect no such correlation and it is instructive to consider how observable errors typically scale in this case. This typical scaling would be revealed by calculating the expected value of the magnitude of the observable error \eqr{eq:error}, averaging over all possible pairings of observable values $O_{\mathbf{i}}$ with errors $P_{\mathbf{i}}- Q_{\mathbf{i}}$. An approximation to this, which we can calculate analytically, is the average over all ways of replacing the set $\{ O_{\bf i} \}$ by a sample of $2^N$ values from itself. This is equivalent to replacing each $O_{\mathbf{i}}$ by a random variable $\hat{O}$ which takes values from the original $\{ O_{\bf i} \}$ with equal probability, and hence its mean and variance, $\mathrm{E}[\hat{O}]$ and $\mathrm{Var}[\hat{O}]$, are the mean and variance of $\{ O_{\bf i} \}$.
This replacement is typically a good approximation for observables like current, density and other correlations. Then, normalizing $\ket{Q}$ such that $\sum_{\mathbf{i}} Q_{\mathbf{i}} = 1$, we make the substitution of $O_{\mathbf{i}}$ for $\hat{O}$ in \eqr{eq:error} and find its expected value and variance to be
\begin{align}
\mathrm{E} \left[  \langle O \rangle^P - \langle O \rangle^Q  \right] &= 0 , \nonumber \\
\mathrm{Var} \left[ \langle O \rangle^P - \langle O \rangle^Q  \right] &=  \mathrm{Var} \left[  \hat{O}  \right]  \; || \ket{P} - \ket{Q} ||_2^2  . \nonumber
\end{align}
It then follows that
\begin{equation}
\mathrm{E} \left[ \left| \langle O \rangle^P - \langle O \rangle^Q \right| \right] \approx \mathrm{SD} \left[  \hat{O}  \right]  || \ket{P}- \ket{Q} ||_2, \label{eq:l2scaling} \nonumber
\end{equation}
where $\mathrm{SD}$ means the standard deviation. This scales with the $L_2$ rather than the $L_1$-norm error, and we conclude that if there is no correlation between observable values and the approximation method then typically the observable error will scale with the $L_2$-norm error. An example of this is shown in \fir{fig:obserror}(a). Note that due to the possibility of negative probabilities the normalization $\sum_{\mathbf{i}} Q_{\mathbf{i}} = 1$ is not always the same as $|| \ket{Q} ||_1 = 1$.

\section{The Trotter and truncation errors}
\label{sec:trotter_trunc_errors}
In approximating the full stochastic evolution $\exp(H t)$ by a product of $n$ time-steps, each an identical stochastic evolution $S\nolinebreak= \nolinebreak(\prod_{\ell=1}^{N-1}S_\ell)(\prod_{\ell=N-1}^{1}S_\ell)$, we quantify the error by the $L_1$-norm of the residual
\begin{eqnarray}
&&\mathcal{E}_{\textrm{ST}} =  \|\left\{\mathrm{e}^{H t}   - \prod_{j=n}^1 S\right\}\ket{P}\|_1, \nonumber \\
&&\leq \|(\mathrm{e}^{H \delta t}-S)\prod_{j=n-1}^1 S\ket{P}\|_1 + \|\mathrm{e}^{H \delta t}(\mathrm{e}^{H \delta t}-S)\prod_{j=n-2}^1 S\ket{P}\|_1 \nonumber \\
&&  \cdots +  \|\prod_{j=n}^3\mathrm{e}^{H \delta t}(\mathrm{e}^{H \delta t}-S)S\ket{P}\|_1 + \|\prod_{j=n}^2\mathrm{e}^{H \delta t}(\mathrm{e}^{H \delta t}-S)\ket{P}\|_1, \nonumber \\
&&\leq n\|\mathrm{e}^{H \delta t}-S\|_1. \nonumber
\end{eqnarray}
Now $\|\mathrm{e}^{H \delta t}-S\|_1 \sim \delta t^3$, while $n = t/\delta t$ so the error scales at worst as $\mathcal{E}_{\textrm{ST}} \sim t\delta t^2$.

For convenience let's relabel the total set of two-site stochastic evolution operators over the whole simulation as $S_KS_{K-1}\cdots S_k \cdots S_2S_1\ket{P}$ where $K = 2nN$ and the index $k$ denotes the position in the sequence. The second type of error is the truncation error incurred when any two-site operator $S$ is applied to some MPS $\ket{Q}$. Rather than being applied exactly, the subsequent factorization and truncation back to an MPS with a maximum inner-dimension of $\chi$ produces an MPS $\ket{Q'}$. This can be thought of as being equivalent to implementing exactly some other, generally non-stochastic, transformation $T$, which depends on $k$, $\ket{Q}$ and $\chi$, such that $\ket{Q'} = T\ket{Q}$. The error in doing this is then quantified by the $L_1$-norm $\epsilon = ||(S - T)\ket{Q}||_1$. Once the sequence has been performed up to $K$ the accumulated approximation is $T_KT_{K-1}\cdots T_1\ket{Q}$. For each step $k$ it is possible to extract from the TEBD algorithm an upper-bound to the $L_1$-norm error $\epsilon_k = ||(S_k-T_k)T_{k-1}\cdots T_1\ket{P}||_1$  caused by truncation after applying $S_k$ to the accumulative approximate state up to point $k-1$. Specifically for the SVD, for each two-site operation we can calculate the $L_2$-norm error by the Eckart Young theorem \eqr{eq:eckartyoung} and thus get an upper-bound to each $L_1$-norm $\epsilon_k$ by $2^{N/2}$ times this value. This information then provides an upper-bound to the total accumulative $L_1$-norm error due to truncation $\mathcal{E}_{\textrm{tr}}$ after $K$ two-site operations since
\begin{eqnarray}
&&\mathcal{E}_{\textrm{tr}} =  \|\left\{\prod_{k=1}^K S_k  - \prod_{k=1}^K T_k\right\}\ket{P}\|_1, \nonumber \\
&&\leq \|(S_K-T_K)\prod_{k=1}^{K-1} T_k\ket{P}\|_1 + \|S_K(S_{K-1}-T_{K-1})\prod_{k=1}^{K-2} T_k\ket{P}\|_1 \nonumber \\
&&  \cdots +  \|\prod_{k=3}^K S_k(S_2-T_2)T_1\ket{P}\|_1 + \|\prod_{k=2}^K S_k(S_1-T_1)\ket{P}\|_1, \nonumber \\
&&\leq \|(S_K-T_K)\prod_{k=1}^{K-1} T_k\ket{P}\|_1 + \|(S_{K-1}-T_{K-1})\prod_{k=1}^{K-2} T_k\ket{P}\|_1 \nonumber \\
&&  \cdots +  \|(S_2-T_2)T_1\ket{P}\|_1 + \|(S_1-T_1)\ket{P}\|_1, \nonumber \\
&&= \sum_{k=1}^K \epsilon_k . \nonumber
\end{eqnarray}
Thus the truncation error $\mathcal{E}_{\textrm{tr}}$ grows at worst additively during the Trotter sequence and can be monitored within the algorithm by computing the sum of the upper-bounds for each two-site operation. This gives us a conservative upper-bound $\mathcal{E}_{\textrm{tr}}^{+} > \mathcal{E}_{\textrm{tr}}$.

\section{Non-negative matrix factorization algorithms}
\label{sec:nmf_algorithms}
The lack of convexity and non-uniqueness of the NMF problem for the three most common cost functions in \eqr{eq:costfuncs} has prompted a variety of algorithms to be proposed~\cite{Berry2007}. These algorithms tackle the minimization problem directly for the required rank $\chi$ using alternating least-squares, conjugate gradients, multiplicative updates or projected gradient descent~\cite{Berry2007,Lee1999,Lin2007}. Here we shall briefly mention some details for the latter two approaches. The projected gradient descent method additively updates the elements of the $W$ and $H$ matrices as
\begin{eqnarray}
W_{\mathbf{i}\mu} &\mapsto& W_{\mathbf{i}\mu} + \eta_W \frac{\partial F}{\partial W_{\mathbf{i}\mu}}, \nonumber \\
(H^T)_{\mu \mathbf{j}} &\mapsto& (H^T)_{\mu \mathbf{j}} + \eta_H \frac{\partial F}{\partial (H^T)_{\mu \mathbf{j}}}, \nonumber
\end{eqnarray}
where $\eta_W$ and $\eta_H$ are appropriately chosen descent step-sizes. For the $L_1$-norm the relevant derivatives are
\begin{eqnarray}
\frac{\partial F}{\partial W_{\mathbf{i}\mu}} &=& -\sum_\mathbf{j} \frac{\Delta_{\mathbf{i}\mathbf{j}}}{|\Delta_{\mathbf{i}\mathbf{j}}|}(H^T)_{\mu \mathbf{j}} \nonumber \\
 \frac{\partial F}{\partial (H^T)_{\mu \mathbf{j}}} &=& -\sum_\mathbf{i} W_{\mathbf{i}\mu}\frac{\Delta_{\mathbf{i}\mathbf{j}}}{|\Delta_{\mathbf{i}\mathbf{j}}|}, \nonumber
\end{eqnarray}
where $\Delta_{\mathbf{i}\mathbf{j}} = (\Theta-WH^T)_{\mathbf{i}\mathbf{j}}$ is an element of the residual matrix. Under these update rules there is no explicit preservation of the non-negativity of $W$ and $H$ so typically a projection of the updated matrices is made to the positive orthant at each descent step by setting any negative elements to zero. Furthermore there is no preservation of the unit $L_1$-norm of $\Theta$ within the approximation $WH^T$ so this too has to be explicitly enforced~\cite{Berry2007}.

Gradient descent can also be formulated for the $L_2$-norm and the KL divergence. However, by making step sizes $\eta_W$ and $\eta_H$ depend on the matrix element being minimized via specially chosen functions of the current matrices $W$ and $H$, so-called multiplicative algorithms can be derived~\cite{Lee1999,Lee2000}. For the KL divergence the updates to be applied are
\begin{eqnarray}
W_{\mathbf{i}\mu} &\mapsto& \frac{W_{\mathbf{i}\mu}}{\sum_\mathbf{j} (H^T)_{\mu \mathbf{j}}}\sum_{\mathbf{j}}\left\{(H^T)_{\mu \mathbf{j}}\frac{\Theta_{\mathbf{i}\mathbf{j}}}{(WH^T)_{\mathbf{i}\mathbf{j}}}\right\}, \nonumber \\
(H^T)_{\mu \mathbf{j}} &\mapsto& \frac{(H^T)_{\mu \mathbf{j}}}{\sum_\mathbf{i} W_{\mathbf{i}\mu}}\sum_{\mathbf{i}}\left\{W_{\mathbf{i}\mu}\frac{\Theta_{\mathbf{i}\mathbf{j}}}{(WH^T)_{\mathbf{i}\mathbf{j}}}\right\}. \nonumber
\end{eqnarray}
Under these transformations the KL divergence is non-increasing and is invariant if and only if $W$ and $H$ are at a stationary point~\cite{Lee2000}.

Overall, we find that the above NMF algorithms are highly sensitive to the initial conditions. This means that to get reasonable results many random restarts are required and considerable fluctuations in the final cost function value are observed. They also often show slow convergence (if at all) and are generally much slower than the SVD. For the most relevant case of the $L_1$-norm cost function we have found that using a random initial matrix driven to a KL divergence solution often provides a much higher quality result than simply applying the gradient descent to random initial matrices. Despite this, however, we have found that, for our task, the SVD is superior to current NMF algorithms not only at moderate truncations but also at identifying when a near exact low-rank solution exists \footnote{We note that for data-mining applications the SVD is typically used to benchmark what rank reduction would be acceptable for a subsequent NMF, precisely because it identifies a suitable rank which contains most of the data.}. This is an essential ingredient of an effective decimation algorithm.


\begin{thebibliography}{99}

\bibitem{Krug1991}
J.  Krug, Phys. Rev. Lett. {\bf 67}, 1882 (1991).

\bibitem{Grinstein1990}
G. Grinstein, D.-H. Lee and S. Sachdev, Phys. Rev. Lett. {\bf 64}, 1927 (1990).

\bibitem{Schadschneider2002}
A. Schadschneider, Physica A \textbf{313}, 153 (2002).

\bibitem{Helbing2001}
D. Helbing, Rev. Mod. Phys. \textbf{73}, 1067 (2001).
  
\bibitem{Chowdhury2000}
D. Chowdhury, L. Santen and A. Schadschneider, Phys. Rep. \textbf{329}, 199 (2000).

\bibitem{Chowdhury2005}
D. Chowdhury, A. Schadschneider and K. Nishinari, Phys. Life Rev. \textbf{2}, 318 (2005).

\bibitem{Aghababaie1999}
Y. Aghababaie, G. I. Menon and M. Plischke, Phys. Rev. E \textbf{59}, 2578 (1999).

\bibitem{Chou1999}
T. Chou and D. Lohse, Phys. Rev. Lett. \textbf{82}, 3552 (1999).

\bibitem{Parmeggiani2003}
A. Parmeggiani, T. Franosch and E. Frey, Phys. Rev. Lett. \textbf{90}, 086601 (2003).

\bibitem{Derrida1993}
B. Derrida, M. R. Evans, V. Hakim and V. Pasquier, J. Phys. A \textbf{26}, 1493 (1993).

\bibitem{Blythe2007}
R. A. Blythe and M. R. Evans, J. Phys. A \textbf{40}, R333 (2007).

\bibitem{Schutz2007}
G. M. Sch\"{u}tz, in {\em Phase Transitions and Critical Phenomena}, edited by C. Domb and J. L. Lebowitz (Academic Press, London, 2000), Vol. 19.

\bibitem{Vidal2003}
G. Vidal, Phys. Rev. Lett. \textbf{91}, 147902 (2003).

\bibitem{Vidal2004}
G. Vidal, Phys. Rev. Lett. \textbf{93}, 040502 (2004).
  
\bibitem{Gobert2005}
D. Gobert, C. Kollath, U. Schollw\"{o}ck and G. Sch\"{u}tz, Phys. Rev. E \textbf{71}, 036102 (2005).

\bibitem{White2004}
S. R. White and A. E. Feiguin, Phys. Rev. Lett. {\bf 93}, 076401 (2004).
  
\bibitem{Clark2004}
S. R. Clark and D. Jaksch, Phys. Rev. A {\bf 70}, 043612 (2004).

\bibitem{Daley2004}
A. J. Daley, C. Kollath, U. Schollw\"{o}ck and G. Vidal, J. Stat. Mech. P04005 (2004).

\bibitem{Jaksch2004}
D. Jaksch, Contemporary Physics {\bf 45}, 367 (2004).

\bibitem{Daley2005}
A. J. Daley, S. R. Clark, D. Jaksch and P. Zoller, Phys. Rev. A {\bf 72}, 043618 (2005).
 
\bibitem{Clark2006}
S. R. Clark and D. Jaksch, New J. Phys. {\bf 8}, 160 (2006).  

\bibitem{Bruderer2007}
M. Bruderer, A. Klein, S. R. Clark and D. Jaksch, Phys. Rev. A {\bf 76}, 011605(R) (2007).  

\bibitem{Bruderer2008}
M. Bruderer, A. Klein, S. R. Clark and D. Jaksch, New J. Phys. 10, 033015 (2008).  

\bibitem{Gillespie1976}
D. T. Gillespie, J. Comput. Phys. {\bf 22}, 403 (1976).

\bibitem{Popkov2008}
V. Popkov, M. Salerno and G. M. Sch\"{u}tz, Phys. Rev. E {\bf 78}, 011122 (2008).

\bibitem{Lipowski2009}
A. Lipowski and D. Lipowska, Phys. Rev. E {\bf 79}, 060102(R) (2009).

\bibitem{Fannes1992}
M. Fannes, B. Nachtergaele, and R. F. Werner, Commun. Math. Phys. \textbf{144}, 443 (1992).
  
  \bibitem{Ostlund1995}
S. \"{O}stlund and S. Rommer, Phys. Rev. Lett. {\bf 75}, 3537 (1995).

\bibitem{Rommer1996}
S. Rommer and S. \"{O}stlund, Phys. Rev. B \textbf{55}, 2164 (1997).

\bibitem{PerezGarcia2007}
D. Perez-Garcia, F. Verstraete, M. M. Wolf and J. I. Cirac, Quant. Inf. Comput. \textbf{7}, 401 (2007).  
  
\bibitem{White1992}
S. R. White, Phys. Rev. Lett. \textbf{69}, 2863 (1992).

\bibitem{White1993}
S. R. White, Phys. Rev. B \textbf{48}, 10345 (1993).

\bibitem{Schollwock2005}
U. Schollw\"{o}ck, Rev. Mod. Phys. \textbf{77}, 259 (2005).

\bibitem{Hieida1998}
Y. Hieida, J. Phys. Soc. Jpn.  \textbf{67}, 369 (1998).

\bibitem{Carlon1999}
E. Carlon, M. Henkel and U. Schollw\"{o}ck, Eur. Phys. J. B \textbf{12}, 99 (1999).

\bibitem{Carlon2001}
E. Carlon, M. Henkel and U. Schollw\"{o}ck, Phys. Rev. E \textbf{63}, 036101 (2001).
  
\bibitem{Glauber1963}
R. J. Glauber, J. Math. Phys. {\bf 4}, 294 (1963).

\bibitem{Alcaraz1993}
F. C. Alcaraz, M. Droz, M. Henkel and V. Rittenberg, Ann. Phys. {\bf 230}, 250 (1994).
  
\bibitem{Golinelli2006}
O. Golinelli and K. Mallick, J. Phys. A {\bf 39}, 12679 (2006).

\bibitem{Derrida1998}
B. Derrida, Phys. Rep. \textbf{301}, 65 (1998).

\bibitem{Schmittmann2007}
B. Schmittmann and R. K. P. Zia, in {\em Phase Transitions and Critical Phenomena}, edited by C. Domb and J. L. Lebowitz (Academic Press, New York, 1995), Vol. 17.

\bibitem{Stinchcombe2001}
R. Stinchcombe, Adv. Phys. \textbf{50}, 431 (2001).

\bibitem{Sachdev2001}
S. Sachdev, {\em Quantum Phase Transitions} (Cambridge University Press, Cambridge, 2001).
  
\bibitem{Rajewsky2004}
N. Rajewsky, L. Santen, A. Schadschneider and M. Schreckenberg, J. Stat. Phys. \textbf{92}, 151 (2004).
  
\bibitem{Golub1996}
G. H. Golub and C. F. V. Loan, {\em Matrix Computations} (Johns Hopkins University Press, Baltimore, 1996), 3rd Ed.

\bibitem{Paatero1994}
P. Paatero and U. Tapper, Environmetrics {\bf 5}, 111 (1994).
 
\bibitem{Lee1999}
D. D. Lee and H. S. Seung, Nature {\bf 401}, 788 (1999).

\bibitem{Temme2010}
K. Temme and F. Verstraete, Phys. Rev. Lett. {\bf 104}, 210502 (2010).   
  
\bibitem{Verstraete2009}
F. Verstraete, V. Murg and J. I. Cirac, Adv. Phys. {\bf 57}, 143 (2008).

\bibitem{Cirac2009}
J. I. Cirac and F. Verstraete, J. Phys. A {\bf 42}, 504004 (2009).

\bibitem{Verstraete2006}
F. Verstraete and J. I. Cirac, Phys. Rev. B {\bf 73}, 094423 (2006).

\bibitem{Nielson2000}
M. A. Nielson and I. L. Chuang, {\em Quantum Computation and Quantum Information} (Cambridge University Press, Cambridge, 2000).
  
\bibitem{Amico2008}
L. Amico, R. Fazio, A. Osterloh and V. Vedral, Rev. Mod. Phys. \textbf{80}, 517 (2008).

\bibitem{Eisert2010}
J. Eisert, M. Cramer and M. B. Plenio, Rev. Mod. Phys. {\bf 82}, 277 (2010). 

\bibitem{Gaussier2005}
E. Gaussier and C. Goutte, in {\em Proceedings of the 28th annual international ACM SIGIR conference on Research and development in information retrieval} (Salvador, 2005), p. 601.

\bibitem{Eckart1936}
C. Eckart and G. Young, Psychometrika {\bf 1}, 211 (1936).

\bibitem{Seneta2006}
E. Seneta, {\em Non-negative matrices and Markov chains} (Springer, New York, 2006), 2nd Ed.

\bibitem{Suzuki1990}
M. Suzuki, Phys. Lett. A \textbf{146}, 319 (1990); J. Math. Phys. \textbf{32}, 400 (1991).

\bibitem{Vasiloglou2009}
N. Vasiloglou, A. G. Gray and D. V. Anderson, in {\em Proceedings of the Ninth SIAM International Conference on Data Mining} (Sparks, Nevada, 2009), p. 673.

\bibitem{Lee2000}
D. D. Lee and H. S. Seung, Advances in Neural Information Processing Systems {\bf 13}, 556 (2000).

\bibitem{VidalPrivate}
M. Zwolak and G. Vidal (private communication, 2004).

\bibitem{Keizer1972}
J. Keizer, J. Stat. Phys. {\bf 6}, 67 (1972).

\bibitem{Shi2006}
Y. Y. Shi, L. M. Duan and G. Vidal, Phys. Rev. A \textbf{74}, 022320 (2006).

\bibitem{Markov2009}
I. L. Markov, Y. Y. Shi, Algorithmica DOI: 10.1007/s00453-009-9312-5 (2009).

\bibitem{Huebener2010}
R. Huebener, V. Nebendahl and W. D\"{u}r, New J. Phys. {\bf 12}, 025004 (2010).

\bibitem{Frowis2010}
F. Fr\"{o}wis, V. Nebendahl and W. D\"{u}r, preprint arXiv:1003.1047v1 (2010).

\bibitem{Banuls2009}
M. C. Ba\~{n}uls, M. B. Hastings, F. Verstraete and J. I. Cirac, Phys. Rev. Lett. {\bf 102}, 240603 (2009).

\bibitem{Stoudenmire2010}
E. M. Stoudenmire and S. R. White, New J. Phys. {\bf 12}, 055026 (2010).

\bibitem{Berry2007}
M. W. Berry, M. Browne, A. N. Langville, V. P. Pauca and R. J. Plemmons, Computational Statistics and Data Analysis  {\bf 52}, 155 (2007).

\bibitem{Lin2007}
C.-J. Lin, Neural Computation {\bf 19}, 2756 (2007).

\end{thebibliography}
\end{document}